\documentclass[aps, prd, floatfix, nofootinbib, showpacs, twocolumn, superscriptaddress]{revtex4-1}

\usepackage{amsmath, amsfonts, amssymb, bm}
\usepackage{graphicx}
\usepackage{color}
\usepackage{subfigure}
\usepackage{multirow}
\usepackage{textcomp}
\usepackage{slashed}

\usepackage{enumitem}

\usepackage[mathscr,scaled=1.15]{urwchancal}
\DeclareFontFamily{OT1}{pzc}{}
\DeclareFontShape{OT1}{pzc}{m}{it}%
{<-> s * [1.15] pzcmi7t}{}
\DeclareMathAlphabet{\mathpzc}{OT1}{pzc}{m}{it}

\definecolor{purple}{rgb}{0.5,0,0.5}
\definecolor{blue}{rgb}{0.0,0,0.9}
\definecolor{prdblue}{rgb}{0.133,0.118,0.498}
\usepackage[colorlinks=true, pdfstartview=FitV, linkcolor=prdblue, citecolor= prdblue, urlcolor=prdblue]{hyperref}

\begin{document}

\title{\mbox{\textbf{$\gamma^\ast \gamma \to \eta, \eta^\prime$}} transition form factors}

\author{Minghui Ding}
\affiliation{School of Physics, Nankai University, Tianjin 300071, China}
\affiliation{Physics Division, Argonne National Laboratory, Argonne, Illinois
60439, USA}
\affiliation{European Centre for Theoretical Studies in Nuclear Physics
and Related Areas (ECT$^\ast$) and Fondazione Bruno Kessler\\ Villa Tambosi, Strada delle Tabarelle 286, I-38123 Villazzano (TN) Italy}

\author{Kh\'epani Raya}
\affiliation{School of Physics, Nankai University, Tianjin 300071, China}
\affiliation{Instituto de F\'{\i}sica y Matem\'aticas, Universidad
Michoacana de San Nicol\'as de Hidalgo\\
Edificio C-3, Ciudad Universitaria, C.P. 58040,
Morelia, Michoac\'an, M{\'e}xico}

\author{\\Adnan Bashir}
\affiliation{Instituto de F\'{\i}sica y Matem\'aticas, Universidad
Michoacana de San Nicol\'as de Hidalgo\\
Edificio C-3, Ciudad Universitaria, C.P. 58040,
Morelia, Michoac\'an, M{\'e}xico}

\author{Daniele Binosi}
\affiliation{European Centre for Theoretical Studies in Nuclear Physics
and Related Areas (ECT$^\ast$) and Fondazione Bruno Kessler\\ Villa Tambosi, Strada delle Tabarelle 286, I-38123 Villazzano (TN) Italy}

\author{Lei Chang}
\email[]{leichang@nankai.edu.cn}
\affiliation{School of Physics, Nankai University, Tianjin 300071, China}

\author{Muyang Chen}
\affiliation{School of Physics, Nankai University, Tianjin 300071, China}

\author{Craig D.~Roberts}
\email[]{cdroberts@anl.gov}
\affiliation{Physics Division, Argonne National Laboratory, Argonne,
Illinois 60439, USA}

\date{29 October 2018}

\begin{abstract}
Using a continuum approach to the hadron bound-state problem, we calculate $\gamma^\ast \gamma \to \eta, \eta^\prime$ transition form factors on the entire domain of spacelike momenta, for comparison with existing experiments and in anticipation of new precision data from next-generation $e^+ e^-$ colliders.  One novel feature is a model for the contribution to the Bethe-Salpeter kernel deriving from the non-Abelian anomaly, an element which is crucial for any computation of $\eta, \eta^\prime$ properties.  The study also delivers predictions for the amplitudes that describe the light- and strange-quark distributions within the $\eta, \eta^\prime$.  Our results compare favourably with available data.  Important to this at large-$Q^2$ is a sound understanding of QCD evolution, which has a visible impact on the $\eta^\prime$ in particular.  Our analysis also provides some insights into the properties of $\eta, \eta^\prime$ mesons and associated observable manifestations of the non-Abelian anomaly.
\end{abstract}

\maketitle

\section{Introduction}
\label{SecIntroduction}
%
Quantum chromodynamics (QCD) describes the strong interaction sector of the Standard Model and its influence on hadron electroweak properties.  Despite having emerged more than forty years ago, from an array of distinct ideas and discoveries \cite{Marciano:1977su, Marciano:1979wa}, there are few predictions for processes that involve strong-QCD dynamics, such as hadron elastic and transition form factors.  The cleanest relate to $\gamma^\ast \gamma^{(\ast)} \to M$ transition form factors, $G_M(Q^2)$, where $M$ is a pseudoscalar meson.

Focusing on $\gamma^\ast \gamma \to M$ and considering any $q\bar q$ component of $M$, then $\exists \, Q_0>\Lambda_{\rm QCD}$ such that \cite{Lepage:1980fj}
\begin{equation}
\label{EqHardScattering}
Q^2 G_M^q(Q^2) \stackrel{Q^2 > Q_0^2}{\approx} 4 \pi^2 \, f_{M}^q\, {\mathpzc e}_q^2\, \tilde{\mathpzc{w}}_{M}^q(Q^2),
\end{equation}
where:
$\Lambda_{\rm QCD} \sim 0.2\,$GeV is the empirical mass-scale of QCD;
$f_M^q$ is the $q\bar q$-component contribution to the pseudovector projection of the meson's wave function onto the origin in configuration space, \emph{i.e}.\ a leptonic decay constant;
${\mathpzc e}_q$ is the electric charge of quark $q$;
and
\begin{equation}
\label{wphi}
\tilde {\mathpzc{w}}_{M}^q(Q^2) = \int_0^1 dx\, \frac{1}{x} \,\varphi_{M}^q(x;Q^2)\,,
\end{equation}
where $\varphi_{M}^q(x;Q^2)$ is the dressed-valence $q$-parton contribution to the meson's distribution amplitude (DA).  %
The DA in Eq.\,\eqref{wphi} is determined by the meson's light-front wave function and relates to the probability that, with constituents collinear up to the scale $\zeta=\surd Q^2$, the dressed-valence $q$-parton carries light-front fraction $x$ of the bound-state's total momentum.
The complete transition form factor is obtained as a sum over the various $q\bar q$ subcomponent contributions:
\begin{equation}
G_M = \sum_{q\in M} \psi_M^q G_M^q,
\end{equation}
where $\psi_M^q$ is a flavour weighting factor originating in the meson's wave function.

Notably \cite{Lepage:1979zb, Efremov:1979qk, Lepage:1980fj} ($\tau^2 := \Lambda_{\rm QCD}^2/Q^2$):
\begin{equation}
\label{PDAcl}
\varphi_M(x;Q^2) \stackrel{\tau \simeq 0}{\approx} \varphi_{\infty}(x) = 6 x (1-x)\,,
\end{equation}
\emph{i.e}.\ the DA acquires its asymptotic profile and hence
\begin{equation}
\label{GPqcl}
Q^2 G_M^q(Q^2) \stackrel{\tau\simeq 0}{\approx} 12 \pi^2 \, f_{M}^q\, {\mathpzc e}_q^2\,.
\end{equation}
Consequently, on $\tau\simeq 0$ the $\gamma^\ast \gamma \to M$  transition form factor exhibits simple scaling; and the anomalous dimension, characteristic of gauge field theories quantised in four dimensions, is ``hidden'' in the manner of approach to the $\tau=0$ limit.  (N.B.\ As will become clear, owing to the non-Abelian axial anomaly in QCD, Eq.\,\eqref{GPqcl} is amended when $M=\eta$, $\eta^\prime$ \cite{Feldmann:1997vc, Agaev:2014wna}.)

An array of experiments have been performed with a view to testing Eqs.\,\eqref{EqHardScattering}, \eqref{GPqcl} for the neutral pion \cite{Behrend:1990sr, Gronberg:1997fj, Aubert:2009mc, Uehara:2012ag}.  Such measurements are difficult, typically involving the study of $e^+$-$e^-$ collisions, in which one of the outgoing fermions is detected after a large-angle scattering whilst the other is scattered through a small angle and, hence, undetected.  The detected fermion is assumed to have emitted a highly-virtual photon, the undetected fermion, a soft-photon; and these photons are supposed to fuse and produce the final-state pseudoscalar meson.  There are many possible background processes and loss mechanisms in this passage of events, and thus ample room for systematic error, especially as $Q^2$ increases \cite{Bevan:2014iga}.

The potential for such errors probably plays a large part in the controversy surrounding the most recent measurements of $\gamma^\ast \gamma \to \pi^0$ \cite{Aubert:2009mc, Uehara:2012ag}, which exhibit incompatible trends in their evolution with photon virtuality \cite{Stefanis:2012yw} and have drawn much attention, \emph{e.g}.\ Refs.\,\cite{Mikhailov:2009kf, Radyushkin:2009zg, Polyakov:2009je, Roberts:2010rn, Agaev:2010aq, Brodsky:2011yv, Balakireva:2011wp, Nedelko:2016vpj, Eichmann:2017wil, Choi:2017zxn}.  In this connection, a theoretical framework that provides a unified treatment of the charged-pion elastic electromagnetic form factor, its valence-quark distribution function and amplitude, and numerous other qualities, was found \cite{Raya:2015gva} to deliver a prediction for $G_\pi(Q^2)$ that agrees with all available data, except that in Ref.\,\cite{Aubert:2009mc}, and is fully consistent with Eq.\,\eqref{EqHardScattering}.  Moreover, it revealed that Eq.\,\eqref{EqHardScattering} provides an accurate representation of the neutral-pion transition form factor on $Q^2\gtrsim 15\,$GeV$^2$.

Experimental data on the $\gamma^\ast \gamma \to \eta_c$ transition are also available \cite{Lees:2010deBaBar}.  In this case, the computational framework used for $G_\pi$ produces a result for $G_{\eta_c}(Q^2)$ which matches the data and is broadly consistent with Eq.\,\eqref{EqHardScattering} on $Q^2\gtrsim 30\,$GeV$^2$ so long as the DA used to describe the $\eta_c$ is that appropriate to the experimental scale, not the asymptotic limit \cite{Raya:2016yuj, Chen:2016bpj}.  The predictions in Refs.\,\cite{Raya:2016yuj, Chen:2016bpj} are confirmed by a recent next-to-next-to-leading (NNLO) order analysis using nonrelativistic QCD (nrQCD) \cite{Wang:2018lry}.

There is no empirical information on the $\gamma^\ast \gamma \to \eta_b$ transition, but the predictions in Refs.\,\cite{Raya:2016yuj, Chen:2016bpj} agree with a NNLO nrQCD analysis \cite{Feng:2015uha}.  They also reveal that, at realistically accessible momentum transfers, owing to the size of the $\eta_b$ mass, Eq.\,\eqref{EqHardScattering} overestimates the direct calculation by a factor of approximately two, even when an $\eta_b$ DA appropriate to the experimental scale is used.

These remarks show that a unified description of the transitions $\gamma^\ast \gamma \to M$, $M=\pi^0$, $\eta_c$, $\eta_b$, is now available along with an understanding of the applicability of Eq.\,\eqref{EqHardScattering} in each case \cite{Raya:2015gva, Raya:2016yuj, Chen:2016bpj}.  Wanting, however, are equivalent explanations of $\gamma^\ast \gamma \to (\eta,\eta^\prime)$.  Importantly, given that relevant data exist on the domain $Q^2\in [0,112]\,$GeV$^2$ \cite{Gronberg:1997fj, BABAR:2011ad, Aubert:2006cy}, then requiring a theoretical framework to unify the description of these transitions with those reviewed above, on the entire domain $Q^2\geq 0$, is a severe test of the approach.  The challenge is compounded by the fact that the flavour structure of the $\eta$, $\eta^\prime$ mesons is a measure of the strength of the non-Abelian anomaly and topological effects within hadrons \cite{Christos:1984tu, Bhagwat:2007ha}.  Hence, those truncations of the continuum two-valence-body bound-state problem which are typically employed cannot provide a realistic description of the $\eta$, $\eta^\prime$ mesons.

Herein, we extend the approach of Refs.\,\cite{Raya:2015gva, Raya:2016yuj}, introducing contributions to the meson Bethe-Salpeter kernels which express effects arising from the non-Abelian anomaly, and deliver predictions for the $\gamma^\ast \gamma \to \eta,\eta^\prime$ transition form factors on $Q^2\geq 0$.  In doing so, we complete a unified description of the two-photon transition form factors of all charge-neutral ground-state pseudoscalar mesons, including a discussion of the relevance of Eq.\,\eqref{EqHardScattering} to understanding each case.
Section~\ref{Sec2} introduces the $\eta, \eta^\prime$-mesons as a continuum bound-state problem, reviewing the issue of flavour mixing, describing the matter-sector equations relevant to mesons, and detailing the kernels used in solving them.
The solutions are discussed in Sec.\,\ref{sec3}, along with predictions for the $\eta, \eta^\prime$ masses and widths, and calculations of the dressed-valence-quark DAs that represent the $\eta, \eta^\prime$-mesons.  (A discussion of the topological charge contained within these systems and detailed descriptions of the perturbation theory integral representations (PTIRs) \cite{Nakanishi:1969ph} used to interpolate the numerical solution arrays are provided in two separate appendices.)
The $\gamma^\ast \gamma \to \eta,\eta^\prime$ transition form factors are reported and analysed in Sec.\ref{sec4}, with particular attention being paid to the impact of QCD evolution on the Bethe-Salpeter wave functions and, consequently, the transition form factors.
Section~\ref{SecEpilogue} provides a summary and perspective.

\section{\mbox{\bf $\eta$}, \mbox{\bf $\eta^\prime$} as two-body bound states}
\label{Sec2}
\subsection{Flavour basis}
\label{SecFlavourBasis}
We consider the limit of perfect isospin symmetry, in which case the $\pi^0$ does not mix with $\eta$, $\eta^\prime$.  As discussed elsewhere \cite{Bhagwat:2007ha}, this is a good approximation: the $s\bar s$ component of the physical $\pi^0$ Bethe-Salpeter amplitude is roughly 1\%, corresponding to a $\pi^0$-$\eta$ mixing angle of $\lesssim 1^\circ$.  These features were exploited in explaining the $\gamma^\ast \gamma \to \pi^0$ transition \cite{Raya:2015gva}.

In discussing $\eta$-$\eta^\prime$ mixing, it is often convenient to work with the $U(N_f=3)$ quark flavour basis \cite{Feldmann:1998vh, Feldmann:1998sh}, in which case the associated Bethe-Salpeter wave functions can be written $(l=u=d)$
\begin{subequations}
\begin{align}
\chi_{\eta,\eta^\prime}&(k;P)  = \mathbb{F}^l \chi^l_{\eta,\eta^\prime}(k;P) + \mathbb{F}^s\chi_{\eta,\eta^\prime}^s(k;P)\,,
\label{chiFB}\\
\mathbb{F}^l & =
\left(\begin{array}{ccc}
1 & 0 & 0 \\
0 & 1 & 0 \\
0 & 0 & 0 \\
\end{array}\right)\,,\quad
\mathbb{F}^s  =
\left(\begin{array}{ccc}
0 & 0 & 0 \\
0 & 0 & 0 \\
0 & 0 & \surd 2 \\
\end{array}\right)\,.
\end{align}
\end{subequations}
The coefficients $\chi^{l,s}_{\eta,\eta^\prime}(k;P)$ in Eq.\,\eqref{chiFB} are Bethe-Salpeter wave functions which, respectively, describe the momentum-space $l\bar l$ or $s \bar s$ correlations in the $\eta$, $\eta^\prime$: $k$ is the relative momentum between the valence-quarks and $P$ is the bound-state's total momentum.

Meson Bethe-Salpeter amplitudes, $\Gamma$, are obtained from the wave functions by amputating the quark legs:
\begin{align}
\chi_{\eta,\eta^\prime}^{l,s}(k;P) & =
S_{l,s}(k_+) \Gamma_{\eta,\eta^\prime}^{l,s}(k;P) S_{l,s}(k_-) \,,
\end{align}
where $k_\pm = k \pm P/2$ and $S_{l,s}$ are dressed-quark propagators.  Defining ${\mathpzc S} = {\rm diag}[S_l,S_l,S_s]$, then
\begin{subequations}
\begin{align}
\chi_{\eta,\eta^\prime}(k;p) & = {\mathpzc S}(k_+) \Gamma_{\eta,\eta^\prime}(k;P)  {\mathpzc S}(k_-)\,,\\
\Gamma_{\eta,\eta^\prime}(k;P)  & = \mathbb{F}^l \Gamma^l_{\eta,\eta^\prime}(k;P) + \mathbb{F}^s\Gamma_{\eta,\eta^\prime}^s(k;P)\,.
\label{GammaFB}
\end{align}
\end{subequations}
For any pseudoscalar meson, or flavour-separated subcomponent, the amplitudes in Eq.\,\eqref{GammaFB} have the form:
\begin{subequations}
\label{EqGenBSA}
\begin{align}
&\Gamma(k;P)  = \sum_{i=1}^4 g_i(k;P)\, {\mathbb D}_i(k;P)\,, \\
%
%
& \begin{array}{llll}
{\mathbb D}_1(k;P)  & =  i \gamma_5\,, & {\mathbb D}_2(k;P) & = \gamma_5\gamma\cdot P\,, \\
{\mathbb D}_3(k;P)  & = \gamma_5 \gamma\cdot k \,, & {\mathbb D}_4(k;P) & = \gamma_5\sigma_{\mu\nu} k_\mu P_\nu \,, \\
\end{array}
\end{align}
\end{subequations}
where $\{g_i(k;P)|i=1,\ldots,4\}$ are scalar functions.

\subsection{Gap equations}
The natural first step in a continuum analysis of the valence-quark+valence-antiquark bound-state problem is computation of the one-body propagators for the fermions involved: in this case, $l=u=d$- and $s$-quarks.  These propagators can be obtained from the associated gap equations:
\begin{subequations}
\label{EqGap}
\begin{align}
S^{-1}_{l,s}(k) & = Z_2 (i \gamma\cdot k + m_{l,s}^{\rm bm}) + \Sigma_{l,s}(k)\,,\\
[\Sigma_{l,s}(k)]_{\iota_1\iota_2} & = \int_{dq}^\Lambda
{\!}_{l,s}{\mathpzc J}_{\;\iota_1 \iota_2}^{\iota_1^\prime \iota_2^\prime}(k,q) [S_{l,s}(q)]_{\iota_1^\prime \iota_2^\prime} \,,
\end{align}
\end{subequations}
where: $Z_{2}$ is the quark wave function renormalisation constant and $m_{l,s}^{\rm bm}$ are the quark bare masses; $\int_{dq}^\Lambda$ represents a symmetry-preserving regularisation of the four-dimensional integral;
and ${\!}_{l,s}{\mathpzc J}_{\;\iota_1 \iota_2}^{\iota_1^\prime \iota_2^\prime}$  is the gap equation's kernel, with the indices describing spinor structure (and colour and flavour, when required).
We employ a mass-independent momentum-subtraction renormalisation scheme throughout, implemented by using the scalar Ward-Green-Takahashi identity \cite{Ward:1950xp, Green:1953te, Takahashi:1957xn} and fixing all renormalisation constants in the chiral limit \cite{Chang:2008ec}, with renormalisation scale $\zeta=2\,$GeV$=:\zeta_2$.

The solutions of Eq.\,\eqref{EqGap} take the following form:
\begin{subequations}
\begin{align}
S_{l,s}(k) &
= -i\gamma\cdot k\,\sigma_V^{l,s}(k^2) + \sigma_S^{l,s}(k^2)\,, \label{SfVS}\\
&
= Z_{l,s}(k^2)/[i\gamma\cdot k + M_{l,s}(k^2)]\,,
\end{align}
\end{subequations}
where $M_{l,s}(k^2)=\sigma_S^{l,s}(k^2)/\sigma_V^{l,s}(k^2) $ is the running mass for the indicated quark.

\subsection{Bethe-Salpeter equation}
\label{SecBSE}
With the propagators in hand, one can obtain the bound-state amplitudes from the Bethe-Salpeter equation $(M=\eta,\eta^\prime)$:
\begin{equation}
[\Gamma_{M}(k;P)]_{\iota_1 \iota_2}  = \int_{dq}^\Lambda
{\mathpzc K}_{\;\iota_1 \iota_2}^{\iota_1^\prime \iota_2^\prime}(k,q;P)
[\chi_{M}(q;P)]_{\iota_1^\prime \iota_2^\prime}\,,
\label{EqBSE}
\end{equation}
where ${\mathpzc K}_{\;\iota_1 \iota_2}^{\iota_1^\prime \iota_2^\prime}(k,q;P)$ is the renormalised, fully-amputated quark-antiquark scattering kernel, which is two-particle irreducible with respect to the external quark-antiquark lines and does not contain quark-antiquark to single gauge boson annihilation diagrams.

Bound-state solutions of Eq.\,\eqref{EqBSE} lie at isolated values of $P^2<0$.  In order to locate them, it is useful to insert an ``eigenvalue'', $\lambda(P^2)$, as a multiplicative factor on the right-hand-side.  The resulting equation has a solution at all values of $P^2$; and the true bound-state solution is obtained at that $P^2=-m_{M}^2$ for which $\lambda(-m_M^2)=1$.

This procedure also has another use.  Namely, in the computation of observables, the canonically normalised Bethe-Salpeter amplitude must be used \cite{Nakanishi:1969ph, LlewellynSmith:1969az}.  For the flavour-mixed systems we consider, this means that the Bethe-Salpeter amplitudes should be rescaled to ensure
\begin{align}
& \left[\frac{d \ln \lambda(P^2)}{dP^2}\right]^{-1}_{P^2=-m_M^2}  = 2 \,{\rm tr}\int_{dk}\big[ \bar\Gamma_M^l(k;-P)
\nonumber \\
&  
\rule{5em}{0ex}\times \chi_M^l(k;P) + \bar\Gamma_M^s(k;-P) \chi_M^s(k;P) \big]\,,
\end{align}
where the trace is over colour and spinor indices.

Using the canonically normalised Bethe-Salpeter amplitudes, the leptonic decay constants in Eq.\,\eqref{EqHardScattering} can be computed:
\begin{align}
\label{eqfM}
f_M^{l,s}P_\mu & = Z_2\, {\rm tr}\int_{dk}^\Lambda \gamma_5\gamma_\mu \chi_{M}^{l,s}(k;P)\,.
\end{align}
These decay constants have been used to define a flavour-mixing angle via \cite{Feldmann:1998vh, Feldmann:1998sh}
\begin{equation}
\left(
\begin{array}{cc}
f_\eta^l & f_\eta^s \\
f_{\eta^\prime}^l & f_{\eta^\prime}^s
\end{array}
\right)
=
\left(
\begin{array}{cc}
f^l \cos \phi & - f^s \sin \phi\\
f^l \sin \phi & \phantom{-} f^s \cos\phi
\end{array}
\right)\,,
%
\label{EqMixingAngles}
\end{equation}
where $f^l$, $f^s$ are some ``ideal'' decay constants, which exist in the absence of flavour mixing.  These quantities are not known \emph{a priori}, but will be determined as part of our analysis.  We expect $f^l \approx f_\pi$, $f^s \approx (2 f_K - f_\pi$), with the latter estimate based on an equal spacing rule \cite{Benic:2014mha, Qin:2018dqp}.

It is also possible to describe $\eta$-$\eta^\prime$ mixing via matrix elements of the $U(3)$ flavour-octet and -singlet axial-vector currents \cite{Leutwyler:1997yr}:
\begin{equation}
\left(
\begin{array}{cc}
f_\eta^8 & f_\eta^0\\
f_{\eta^\prime}^8 & f_{\eta^\prime}^0\\
\end{array}\right)
=
\left(\begin{array}{cc}
f_8 \cos\theta_8 & -f_0\sin\theta_0\\
f_8 \sin\theta_8 & \phantom{-}f_0\cos\theta_0\\
\end{array}\right).
\label{OSmixing}
\end{equation}
The octet axial-vector current satisfies a standard Ward-Green-Takahashi identity so $f_8$ in Eq.\,\eqref{OSmixing} is independent of the renormalisation scale, $\zeta$.  On the other hand, the flavour-singlet axial-vector current is anomalous in QCD.  Consequently, the singlet decay constant, $f_0$, and thus $f^{l,s}_{\eta,\eta^\prime}$ in Eq.\,\eqref{eqfM} depend on $\zeta$.  Their decrease with $\zeta$ is not too rapid, however, because the leading contribution to the anomalous dimension is O$(\alpha_S^2)$, where $\alpha_S$ is the QCD running coupling \cite{Leutwyler:1997yr, Feldmann:1997vc, Feldmann:1998sh, Agaev:2014wna}.  Additionally, evolution of the $\eta$, $\eta^\prime$ DAs is more complicated than usual: operator mixing plays a role even at leading order.  As will become apparent, these effects impact strongly upon the $\eta^\prime$ because $\theta_0 \simeq 0$.

\subsection{Kernels for the bound-state equations}
\subsubsection{General observations}
A tractable system of bound-state equations is only obtained once a truncation scheme is specified.  A symmetry-preserving approach is described elsewhere \cite{Munczek:1994zz, Bender:1996bb, Binosi:2016rxz}.  The leading-order term is the rainbow-ladder (RL) truncation.  It is known to be accurate for ground-state light-quark vector- and isospin-nonzero-pseudoscalar-mesons, and related ground-state octet and decouplet baryons \cite{Roberts:2000aa, Maris:2003vk, Chang:2011vu, Bashir:2012fs, Roberts:2015lja, Horn:2016rip, Eichmann:2016yit, Eichmann:2008ef, Eichmann:2012zz} because corrections largely cancel in these channels owing to the preservation of the normal Ward-Green-Takahashi identities ensured by the scheme \cite{Munczek:1994zz, Bender:1996bb, Binosi:2016rxz}.  As noted above, however, the RL truncation and most known improvements thereof \cite{Chang:2009zb, Fischer:2009jm, Williams:2015cvx, Binosi:2016rxz} fail for the $\eta$- and $\eta^\prime$-mesons because they do not produce vertices that satisfy the anomalous axial-vector Ward-Green-Takahashi identities described in Ref.\,\cite{Bhagwat:2007ha}.  Consequently, they lead to ideal mixing in the $\eta$-$\eta^\prime$ sector, represented by $\phi= 0$ in Eq.\,\eqref{EqMixingAngles}, in which case one has the unphysical results $\eta \sim u\bar{u} + d\bar{d}$ and $\eta^\prime \sim s\bar{s}$.

\begin{figure}[t]

\includegraphics[clip,width=0.32\textwidth]{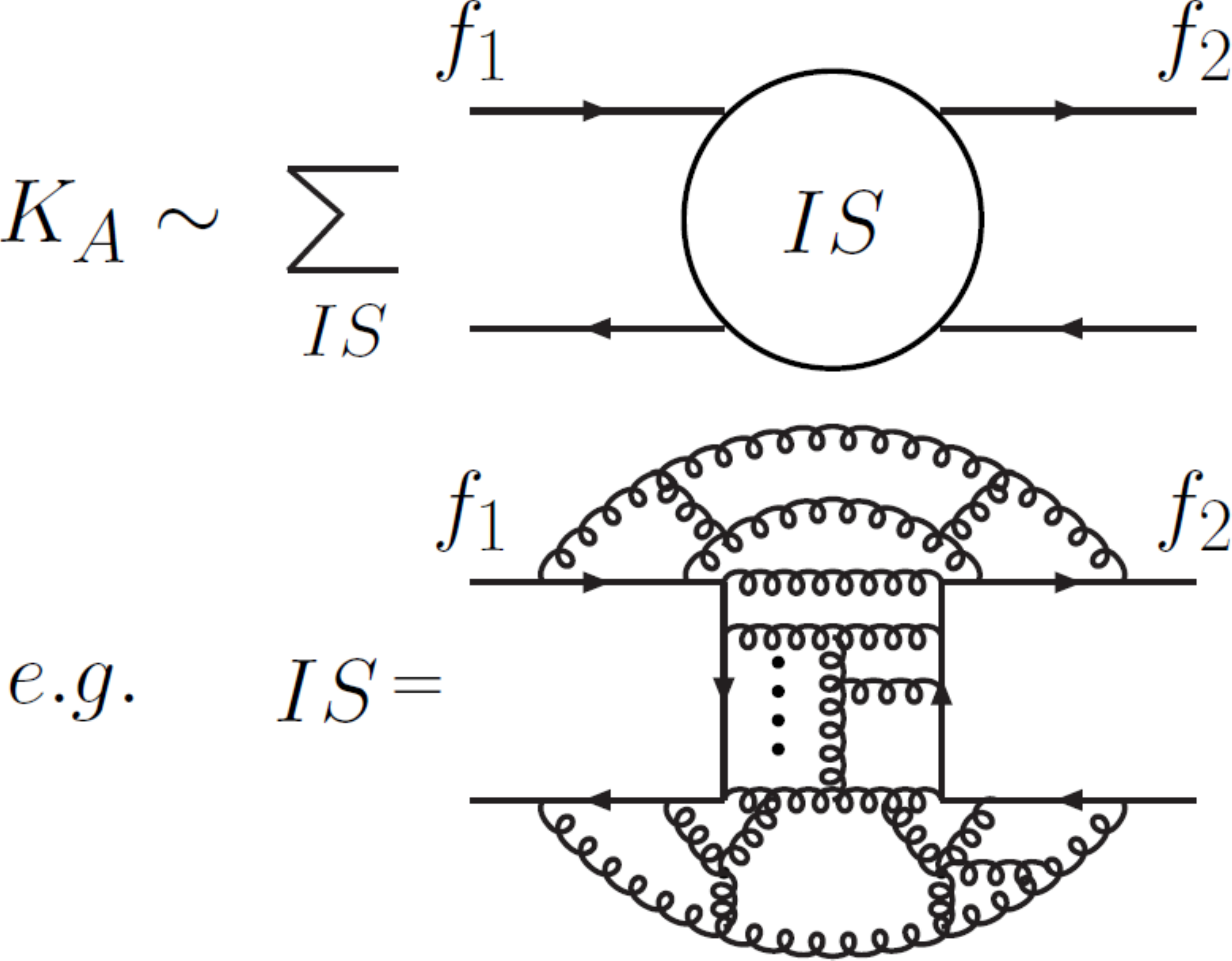}

\caption{\label{FigAnomaly} Any contribution to the Bethe-Salpeter kernel deriving from the non-Abelian anomaly must have the ``hairpin'' structure depicted here; and any intermediate state (\emph{IS}) must involve infinitely many lines.  (Straight lines denote quarks, with $f_1$ and $f_2$ independent, and springs denote gluons.)}
\end{figure}

Considering the structure of the non-Abelian anomaly, it readily becomes apparent that no related contribution to the Bethe-Salpeter kernel can contain external quark or antiquark lines which are simply connected to the internal lines: purely gluonic configurations must mediate, as illustrated in Fig.\,\ref{FigAnomaly}.  Moreover, no finite sum of diagrams can be sufficient.  To understand this remark, focus on any one such single contribution in the chiral limit.  It is necessarily proportional to the total momentum and hence vanishes for $P=0$, thus violating the anomalous Ward-Green-Takahashi identity.  Some coherent effect is required to produce a nonzero contribution.  (As described elsewhere \cite{Christos:1984tu, Bhagwat:2007ha}, variants of the Kogut-Susskind mechanism will suffice \cite{Kogut:1974kt}.)

Following this discussion, it is evident that the Bethe-Salpeter kernel may be decomposed into a sum:
\begin{equation}
\label{EqGenK}
{\mathpzc K} = {\mathpzc K}_N + {\mathpzc K}_A\,,
\end{equation}
where: ${\mathpzc K}_N$ is that part which can readily be constructed diagrammatically and involves all those contributions that are accessible in perturbation theory and therefore contribute at ultraviolet momenta; and ${\mathpzc K}_A$ is the non-Abelian anomaly contribution, depicted in Fig.\,\ref{FigAnomaly}, which is essentially nonperturbative and hence possesses material support only at infrared momenta.

\subsubsection{Rainbow-ladder kernel}
The analyses that provided a unified description of $\gamma^\ast \gamma \to \pi^0, \eta_c, \eta_b$ \cite{Raya:2015gva, Raya:2016yuj} used RL truncation for the gap and Bethe-Salpeter equations, \emph{i.e}.\
\begin{subequations}
\label{KDinteraction}
\begin{align}
{\!}_{l,s}{\mathpzc J}_{\;\iota_1 \iota_2}^{\iota_1^\prime \iota_2^\prime}(k,q) & =
-[\mathscr{K}_N(k,q;P)]_{\iota_1\iota_2}^{\iota_1^\prime \iota_2'} \,,\\
[\mathscr{K}_N(k,q;P)]_{\iota_1\iota_2}^{\iota_1^\prime \iota_2'} & =
\tfrac{4}{3}{\mathpzc G}_{\mu\nu}(t) [i\gamma_\mu]_{\iota_1\iota_1'} [i\gamma_\nu]_{\iota_2^\prime\iota_2}\,,\\
 {\mathpzc G}_{\mu\nu}(t=k-q) & = \tilde{\mathpzc G}(t^2) T_{\mu\nu}(t)\,,
\end{align}
\end{subequations}
where $t^2 T_{\mu\nu}= t^2 \delta_{\mu\nu} - t_\mu t_\nu$.  In this case, Eqs.\,\eqref{EqGap}, \eqref{EqBSE}, as appropriate to $\pi^0$, $\eta_c$, $\eta_b$, could then be solved once $\tilde{\mathpzc G}$ was specified.  Capitalising on two decades of study, Refs.\,\cite{Raya:2015gva, Raya:2016yuj} used the following form \cite{Qin:2011dd, Qin:2011xq} ($s=t^2$):
\begin{align}
\label{defcalG}
 \tfrac{1}{Z_2^2}\tilde{\mathpzc G}(s) & =
 \frac{8\pi^2}{\omega^4} D e^{-\tfrac{s}{\omega^2}} + \frac{8\pi^2 \gamma_m \mathcal{E}(s)}{\ln\big[ {\mathpzc t}+(1+s/\Lambda_{\rm QCD}^2)^2 \big]}\,,
\end{align}
where: $\gamma_m=4/\beta_0=12/(33-2N_f)$, $N_f=4$; $\Lambda_{\rm QCD}=0.234\,$GeV; ${\mathpzc t}={\rm e}^2-1$; and $\mathcal{E}(s) = \{1 - \exp(-s/[4 m_t^2])\}/s$, $m_t=0.5\,$GeV.  (At scales $\zeta>\zeta_2$, all such truncations receive corrections, which typically serve to alter the anomalous dimensions of scale-dependent quantities \cite{Lepage:1980fj}.)

The development of Eqs.\,\eqref{KDinteraction}, \eqref{defcalG} is summarised in Ref.\,\cite{Qin:2011dd}.  Their connection with QCD is described in Ref.\,\cite{Binosi:2014aea}, but it is worth reiterating some points..

The interaction in Eqs.\,\eqref{KDinteraction}, \eqref{defcalG} is deliberately consistent with that determined in studies of QCD's gauge sector, which indicate that the gluon propagator is a bounded, regular function of spacelike momenta that achieves its maximum value on this domain at $k^2=0$ \cite{Bowman:2004jm, Boucaud:2006if, Boucaud:2011ug, Ayala:2012pb, Aguilar:2012rz, Binosi:2014aea, Binosi:2016wcx, Binosi:2016xxu, Binosi:2016nme, Gao:2017uox}, and the dressed-quark-gluon vertex does not possess any structure which can qualitatively alter these features \cite{Skullerud:2003qu, Bhagwat:2004kj, Aguilar:2014lha, Bermudez:2017bpx, Cyrol:2017ewj, Aguilar:2018epe, Oliveira:2018fkj}.
It is specified in Landau gauge because, \emph{e.g}.\ this gauge is a fixed point of the renormalisation group and ensures that sensitivity to differences between \emph{Ans\"atze} for the gluon-quark vertex are least noticeable, thus providing the conditions for which rainbow-ladder truncation is most accurate.  
The interaction also preserves the one-loop renormalisation group behaviour of QCD so that, \emph{e.g}.\ the quark mass-functions produced are independent of the renormalisation point.
On the other hand, in the infrared, \emph{i.e}.\ $k^2 \lesssim (4\Lambda_{\rm QCD})^2$, Eq.\,\eqref{defcalG} defines a two-parameter model, the details of which determine whether confinement and/or dynamical chiral symmetry breaking (DCSB) are realised in solutions of the dressed-quark gap equations.

Computations \cite{Qin:2011dd, Qin:2011xq} reveal that observable properties of light-quark ground-state vector- and isospin-nonzero pseudoscalar-mesons are practically insensitive to variations of $\omega \in [0.4,0.6]\,$GeV, so long as
\begin{equation}
 \varsigma^3 := D\omega = {\rm constant}.
\label{Dwconstant}
\end{equation}
This feature also extends to numerous properties of the nucleon and $\Delta$-baryon \cite{Eichmann:2008ef, Eichmann:2012zz}.  The value of $\varsigma$ is chosen so as to obtain the measured value of the pion's leptonic decay constant, $f_\pi$; and in RL truncation this requires
\begin{equation}
\label{varsigmalight}
\varsigma  =0.80\,{\rm GeV.}
\end{equation}
Refs.\,\cite{Raya:2015gva, Raya:2016yuj} employed $\omega=0.5\,$GeV, the midpoint of the domain of insensitivity.

Given the success of Refs.\,\cite{Raya:2015gva, Raya:2016yuj}, and many other studies that have used the RL truncation to predict hadron observables \cite{Roberts:2000aa, Maris:2003vk, Chang:2011vu, Bashir:2012fs, Roberts:2015lja, Horn:2016rip, Eichmann:2016yit}, we use Eqs.\,\eqref{KDinteraction}, \eqref{defcalG} to define ${\mathpzc K}_N$ in Eq.\,\eqref{EqGenK}.  Implicit in this step is an assumption (made by every practitioner) that all contributions from the non-Abelian anomaly to the dressed-quark gap equation can be -- and are -- absorbed into the value of $\varsigma$; and, hence, that ${\mathpzc K}_A$ in Eq.\,\eqref{EqGenK} describes only those interactions which are essentially four-body in character and therefore cannot be recast as regular corrections to the gluon propagator or gluon-quark vertex.

\subsubsection{Kernel representing the non-Abelian anomaly}
Whilst the RL kernel is constrained by a large body of successful phenomenology, the form of ${\mathpzc K}_A$ is unknown.  On general grounds, given Eq.\,\eqref{EqGenBSA}, its contribution to the Bethe-Salpeter equation for pseudoscalar mesons must take the following form:
\begin{align}
[{\mathpzc K}_A(k,q;P)&]_{\iota_1\iota_2}^{\iota_1^\prime \iota_2^\prime} =\sum_{i=1}^4 \sum_{{\mathsf f}=l,s} {\mathpzc a}_i^{\mathsf f}(k,q;P)
\nonumber \\
&  \times  [{\mathbb F}^{\mathsf f} {\mathbb D}_i(q;P)]_{\iota_2^\prime \iota_1^\prime} [{\mathbb F}^{\mathsf f} {\mathbb D}_i(k;P)]_{\iota_1 \iota_2}\,,
\end{align}
where $\{{\mathpzc a}_i^{\mathsf f}(k,q;P)|i=1,\ldots,4; {\mathsf f} =l,s\}$ are scalar functions, which a detailed analysis of the non-Abelian anomaly could reveal.  That, however, is an independent project.  Required here is just a reasonable model for ${\mathpzc K}_A$, one that produces realistic masses and decay constants for the $\eta$, $\eta^\prime$, because such quantities are the primary impacts of ${\mathpzc K}_A$ to which the spacelike behaviour of the $\gamma^\ast\gamma \to \eta,\eta^\prime$ transition form factors are sensitive.

Following Ref.\,\cite{Bhagwat:2007ha}, we proceed by writing
\begin{subequations}
\label{KAD1}
\begin{align}
\mbox{$\sum_{\mathsf f}$} {\mathpzc a}_1^{\mathsf f} [{\mathbb F}^{\mathsf f}{\mathbb D}_1] & [{\mathbb F}^{\mathsf f}{\mathbb D}_1]  \nonumber \\
& = \xi(s)\cos^2\theta_\xi  [{\mathpzc z}{\mathbb D}_1]_{\iota_2^\prime \iota_1^\prime} [{\mathpzc z}{\mathbb D}_1]_{\iota_1 \iota_2}\,,\\
\mbox{$\sum_{\mathsf f}$} {\mathpzc a}_2^{\mathsf f} [{\mathbb F}^{\mathsf f}{\mathbb D}_2] &  [{\mathbb F}^{\mathsf f}{\mathbb D}_2]  \nonumber \\
& = \frac{1}{{\mathpzc x}^2} \xi(s) \sin^2\theta_\xi
[{\mathpzc z}{\mathbb D}_2]_{\iota_2^\prime \iota_1^\prime} [{\mathpzc z}{\mathbb D}_2]_{\iota_1 \iota_2}\,,
\end{align}
\end{subequations}
${\mathpzc a}_{3,4}^{l,s}= 0$,
where:
${\mathpzc x}= M_{l}(k^2=0)$ is a computed renormalisation-group-invariant mass-scale, characteristic of DCSB;
the parameter $\theta_\xi$ controls the relative strength of the chosen tensor structures;
${\mathpzc z}={\rm diag}[1,1,{\mathpzc r}_A]$, with ${\mathpzc r}_A$ a parameter, introduces a dependence on $U(3)$ flavour-symmetry breaking which models that arising from the dressed-quark lines which complete a ``U-turn'' in the hairpin diagram in Fig.\,\ref{FigAnomaly}; and
%
\begin{equation}
\label{KAD2}
\xi(s) = \frac{8\pi^2}{\omega_\xi^4} D_\xi {\rm e}^{-s/\omega_\xi^2}\,,
\end{equation}
with parameters $D_\xi$, $\omega_\xi$, provides a momentum-dependent interaction strength for the anomaly contribution whose support is localised in the infrared.

\section{\mbox{\bf $\eta$}, \mbox{\bf $\eta^\prime$} Computed Structural Properties}
\label{sec3}
\subsection{Masses and widths}
\label{MassesWidths}
Using the RL truncation parameters described in connection with Eq.\,\eqref{varsigmalight} to define ${\mathpzc K}_{N}$ in Eq.\,\eqref{EqGenK}, and with
\begin{equation}
\hat m_l =   7\,{\rm MeV},\;
\hat m_s = 170\,{\rm MeV}\,,
\end{equation}
which correspond to evolved current-quark masses
\begin{equation}
\label{z2cqmasses}
m_l^{\zeta_2} = 5.1\,{\rm MeV},\; m_s^{\zeta_2} = 125\,{\rm MeV},
\end{equation}
the solution of the relevant coupled gap and Bethe-Salpeter equations yields (in GeV):
\begin{equation}
m_\pi=0.134\,,\; f_\pi=0.093\,,\; m_K=0.496\,,\; f_K=0.11\,,
\end{equation}
in good agreement with experiment \cite{Tanabashi:2018oca}, and the value of ${\mathpzc x}_{\,l}$ listed in Table~\ref{TableA}.

\begin{table}[t]
\caption{\label{TableA}
Parameters appearing in Eqs.\,\eqref{KAD1}, \eqref{KAD2}, which specify the non-Abelian anomaly contribution to our Bethe-Salpeter kernel for $\eta$, $\eta^\prime$ mesons.
(Dimensioned quantities listed in GeV.)}
\begin{center}
\begin{tabular*}
{\hsize}
{
c@{\extracolsep{0ptplus1fil}}
|c@{\extracolsep{0ptplus1fil}}
c@{\extracolsep{0ptplus1fil}}
c@{\extracolsep{0ptplus1fil}}
c@{\extracolsep{0ptplus1fil}}
}\hline
 ${\mathpzc x}_{\,l}\phantom{0}$ & $\surd D_\xi$ & $\omega_\xi$ & $\cos^2\theta_\xi$ & ${\mathpzc r}_A$ \\\hline
$0.51\phantom{0}$ & $0.32$ & $0.30$ & $0.80$ & $0.57$ \\\hline
\end{tabular*}
\end{center}
\end{table}

\begin{table}[t]
\caption{\label{TableB}
Solving the coupled gap and Bethe-Salpeter equations for the $\eta$, $\eta^\prime$ mesons using the parameters described in connection with Table~\ref{TableA}, we find $m_\eta=0.56$, $m_{\eta^\prime}=0.96$ \emph{cf}.\ experiment \cite{Tanabashi:2018oca}: $0.55$, $0.96$, respectively; and the decay constants listed in Row~1.
Row~2 -- Single mixing-angle fit to the results in Row~1, discussed in connection with Eq.\,\eqref{MixingFit}.
Row~3 -- Estimates based on a sample of phenomenological analyses.
(All quantities in GeV.)}
\begin{center}
\begin{tabular*}
{\hsize}
{
l@{\extracolsep{0ptplus1fil}}
|c@{\extracolsep{0ptplus1fil}}
c@{\extracolsep{0ptplus1fil}}
c@{\extracolsep{0ptplus1fil}}
c@{\extracolsep{0ptplus1fil}}
}\hline
 & $f_\eta^l$ & $f_\eta^s$ & $f_{\eta^\prime}^l$ & $f_{\eta^\prime}^s$ \\\hline
herein - direct  & $0.072\phantom{(13)}$ & $-0.092\phantom{(28)}$ & $0.070\phantom{(14)}$ & $0.104\phantom{(8)}$ \\\hline
herein - Eq.\,\eqref{MixingFit}  & $0.074\phantom{(13)}$ & $-0.094\phantom{(28)}$ & $0.068\phantom{(14)}$ & $0.101\phantom{(8)}$ \\\hline
phen.\ \mbox{\cite{Feldmann:1998sh, Benayoun:1999au, DeFazio:2000my}\;}  & $0.090(13)$ & $-0.093(28)$ & $0.073(14)$ & $0.094(8)$\\\hline
\end{tabular*}
\end{center}
\end{table}

We choose the ${\mathpzc K}_A$ parameters, appearing in Eqs.\,\eqref{KAD1}, \eqref{KAD2}, by requiring a fair description of $m_{\eta,\eta^\prime}$, $f_{\eta,\eta^\prime}^{l,s}$: greatest weight is given to the masses in this procedure because they are subject to little uncertainty.  The values listed in Table~\ref{TableA} deliver the results in Table~\ref{TableB}.  For future reference, we highlight that $f_\eta^s$ shows the greatest variation amongst the various estimates, \emph{viz}.\ it is the least well determined by phenomenology.

\begin{figure}[t]

\includegraphics[clip,width=0.45\textwidth]{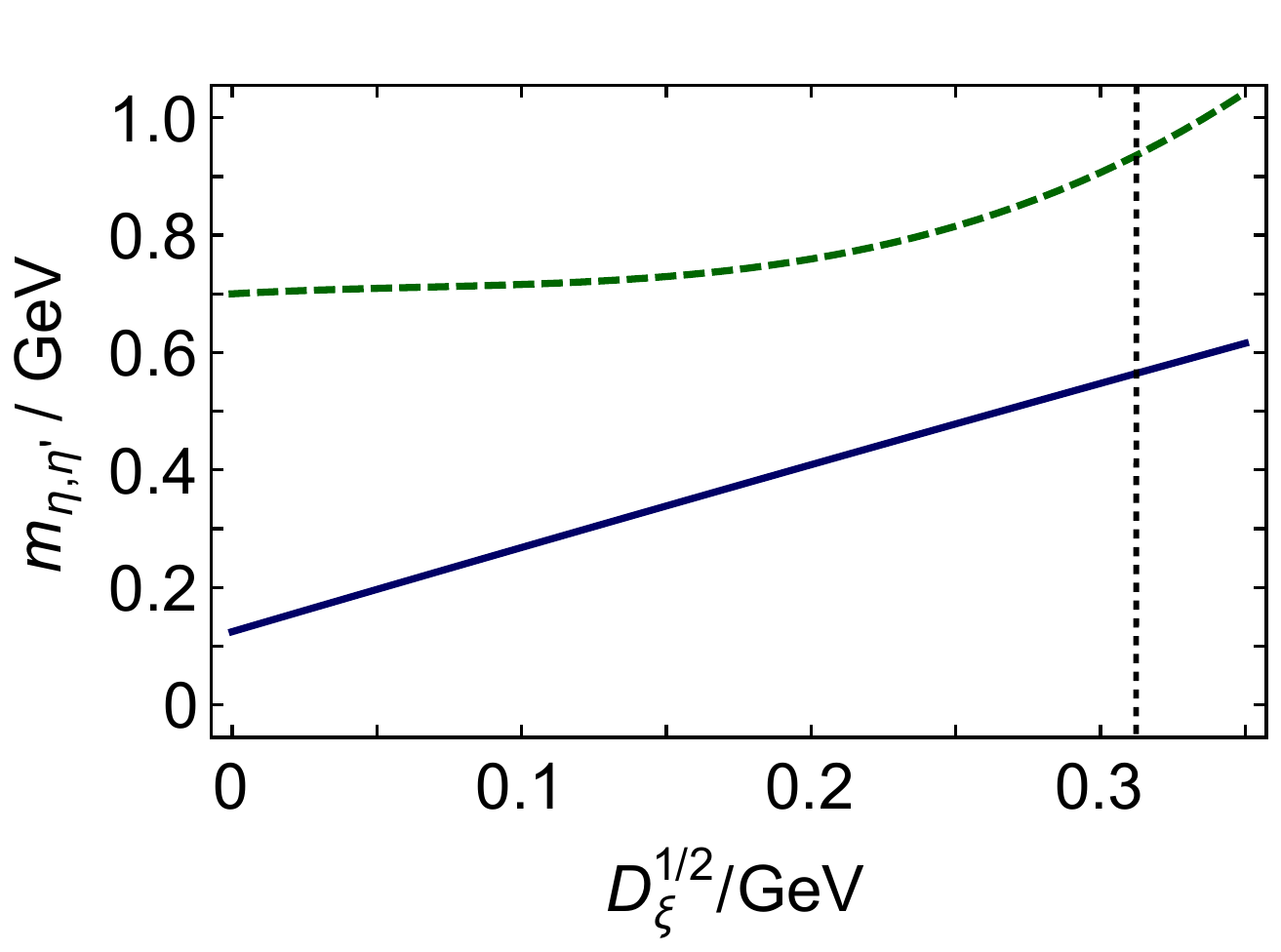}

\caption{\label{FigMassesxi} Growth of $m_\eta$ (solid blue curve) and $m_{\eta^\prime}$ (dashed green curve) with the anomaly-strength parameter in Eq.\,\eqref{KAD2}.  The vertical dotted line marks $\surd D_\xi=0.32\,$GeV, the value which generates are best description of $m_{\eta,\eta^\prime}$.}
\end{figure}

Figure~\ref{FigMassesxi} depicts the evolution of $m_{\eta,\eta^\prime}$ with $\surd D_\xi$ when all other entries in Table~\ref{TableA} are held fixed.  The meson masses evolve smoothly with the anomaly-strength parameter in Eq.\,\eqref{KAD2}: in the absence of an anomaly contribution one has ideal mixing, with $m_\eta = m_\pi$, $m_{\eta^\prime} = m_{s\bar s}=0.7\,$GeV; and both masses grow uniformly with the mixing strength until the empirical values are reached.

We now return to Eq.\,\eqref{EqMixingAngles} and address the question: is there a single mixing angle and pair of ideal decay constants that fairly describe the results in Table~\ref{TableB}?  In answer we report that the values
\begin{subequations}
\label{MixingFit}
\begin{align}
\phi & = 42.8^\circ, \\
f^l & = 0.101\,{\rm GeV} = 1.08\,f_\pi, \\
f^s & =0.138\,{\rm GeV} = 1.49\,f_\pi,
\end{align}
\end{subequations}
yield Row~2 in Table~\ref{TableB}, reproducing our computed results with a root-mean-square difference of 2.4\%.  Notably, the computed values of the ideal decay constants match expectations: $f^l \approx f_\pi$ and $f^s \approx (2 f_K - f_\pi)\approx 1.4 f_\pi$.

The results in Table~\ref{TableB} can readily be translated into values associated with the octet-singlet basis, Eq.\,\eqref{OSmixing}:
\begin{equation}
\label{OSfvalues}
\begin{array}{ll}
f_8 = 1.34\,f_\pi, & \theta_8 = -21^\circ, \\
f_0 = 1.26\,f_\pi, & \theta_0 = -2.8^\circ.
\end{array}
\end{equation}
%
The small value of $\theta_0$ entails that the $\eta$ is largely a flavour-octet state whereas the $\eta^\prime$ is primarily flavour-singlet \cite{Michael:2013gka}.
%
(We discuss the topological charge content of these systems in Appendix~\ref{AppendixTopology}.)
To provide a comparison, we report estimates based on a sample of phenomenological analyses \cite{Feldmann:1998sh, Benayoun:1999au, DeFazio:2000my}:
$f_8 = 1.34(8)\,f_\pi$, $f_0 = 1.25(10)\,f_\pi$;
$\theta_8 = -18(6)^\circ$, $\theta_0 = -6(6)^\circ$.

Adapting current algebra to the present case, one can derive expressions for the $\eta,\eta^\prime \to \gamma\gamma$ decay widths \cite{Feldmann:1997vc}:
\begin{subequations}
\label{eqWidths}
\begin{align}
\Gamma_{\eta\to\gamma\gamma} & = \frac{9 \alpha_{\rm em}^2}{64\pi^3}m_\eta^3
\left[ c_l \frac{f_\eta^l}{(f^l)^2} + c_s \frac{f_\eta^s}{(f^s)^2}\right]^2, \label{eqWidthseta}\\
\Gamma_{\eta^\prime\to\gamma\gamma} & = \frac{9 \alpha_{\rm em}^2}{64\pi^3}m_{\eta^\prime}^3
\left[ c_l \frac{f_{\eta^\prime}^l}{(f^l)^2} + c_s \frac{f_{\eta^\prime}^s}{(f^s)^2}\right]^2,
\end{align}
\end{subequations}
where $c_l=5/9$, $c_s = \surd 2/9$, and $\alpha_{em}\approx 1/137$ is the QED coupling.  These formulae are valid at the resolving scale that defines the computation (in our case, $\zeta_2$); but owing to the scale dependence of $f_0$, as one evolves to a new, larger scale, they receive corrections which ensure that the observable widths are scale independent \cite{Leutwyler:1997yr, Kaiser:2000gs, Feldmann:1999uf}.  Using Eqs.\,\eqref{MixingFit} and the values in Table~\ref{TableB}, Eqs.\,\eqref{eqWidths} yield:
\begin{equation}
\label{EqWidths}
\Gamma_{\eta\to\gamma\gamma}  = 0.42\,{\rm keV}\,,\;
\Gamma_{\eta^\prime\to\gamma\gamma} = 4.66\,{\rm keV}\,,
\end{equation}
predictions which are commensurate with the empirical values, respectively \cite{Tanabashi:2018oca}: $0.516(22)\,$keV, $4.35(36)\,$keV.

\subsection{Integral Representations}
\label{Interpolations}
We now wish to calculate the $\eta$, $\eta^\prime$ leading-twist dressed-valence-quark DAs and the integrals which define our approximation to the $\gamma^\ast\gamma\to \eta, \eta^\prime$ transition form factors, Eq.\,\eqref{TransitionFFs} below.  In computing the quantities discussed in Sec.\,\ref{MassesWidths}, we worked directly with the matrix-valued solutions of the gap and Bethe-Salpeter equations stored simply as large arrays of numbers.  Experience has shown that such input is inadequate for the calculation of DAs and form factors on $Q^2\gtrsim 4\,$GeV$^2$.  We therefore adopt the methods introduced in Refs.\,\cite{Chang:2013pq, Chang:2013nia} and exploited in Refs.\,\cite{Raya:2015gva, Raya:2016yuj}.  Namely, we employ algebraic parametrisations of each array to serve as interpolations.  They are detailed in Appendix~\ref{AppendixA}.

\subsection{Distribution amplitudes}
\label{Distributions}
The DA that describes the light-front longitudinal-momentum distribution for the dressed-quark/-antiquark in a given meson can be obtained by projecting that system's Bethe-Salpeter amplitude onto the light-front, with an appropriate flavour projection.  Herein, therefore, we focus on the following expressions:
\begin{align}
\label{MesonPDA}
f_{M}^{\mathsf f} \varphi_M^{\mathsf f}(x) = Z_2 {\rm tr} \int_{dk}^\Lambda\, \delta_n^x(k_+) \,
\gamma_5\gamma\cdot n \, \chi_M^{\mathsf f}(k;P)\,,
\end{align}
where $\delta_n^x(k_+) = \delta(n\cdot k_+ - x n\cdot P)$, $n^2=0$, $n\cdot P = -m_{M}$, and $f_{M}^{\,{\mathsf f}}$ is the relevant decay constant from Table~\ref{TableB}, Row\,1.

Beginning with Eq.\,\eqref{MesonPDA}, it is straightforward to use the method introduced in Ref.\,\cite{Chang:2013pq} and determine the $\eta$, $\eta^\prime$ DAs by reconstruction from their Mellin moments.  Namely, for each $\varphi_{\eta,\eta^\prime}^{l,s}(x)$, we compute $({\mathpzc y}=2 x-1)$
\begin{subequations}
\begin{align}
& \langle  {\mathpzc y}^m \rangle_M^{\mathsf f} =\int^1_{0} dx\, {\mathpzc y}^{m} \,\varphi_M^{\mathsf f}(x)\\
&=\frac{1}{f_M^{\,{\mathsf f}}}{\rm tr}
Z_2\int^\Lambda_{dk}\frac{(2n{\cdot}k)^m}{(n{\cdot}P)^{m+1}}
\gamma_5\gamma{\cdot}n\chi_M^{\,{\mathsf f}}(k;P)\,. \label{mellinmom}
\end{align}
\end{subequations}
Using the interpolations detailed in Appendix~\ref{AppendixA}, one can readily obtain any finite number of Mellin moments.  We typically use $m_{\rm max}=50$.
Now, since Gegenbauer polynomials of order $\alpha_+=\alpha+1/2$, $\{C_n^{\alpha_+}({\mathpzc y})| n=0,\ldots,\infty\}$, are a complete orthonormal set on ${\mathpzc y}\in[-1,1]$ with respect to the measure $[(1-{\mathpzc y}^2)/4]^{\alpha}$, they enable reconstruction of any function that vanishes at ${\mathpzc y}=-1,1$.
(N.B.\, The DAs we consider are even under ${\mathpzc y}\to -{\mathpzc y}$ and vanish at the endpoints.)
Hence, we write
\begin{equation}
\label{PDAEquation}
\phi^{G_s}(x) = {\mathpzc n}_\alpha \, [(1-{\mathpzc y}^2)/4]^{\alpha} \!
\sum_{j=0,2,\ldots}^{j_s} a_j^\alpha C_j^{\alpha_+}({\mathpzc y})\,,
\end{equation}
${\mathpzc n}_\alpha = \Gamma(2 +2 \alpha)/\Gamma(1+\alpha)^2$, $a_0^\alpha=1$, and minimise
\begin{equation}
\varepsilon_s = \sum_{m=1,\ldots,m_{\rm max}} |\langle {\mathpzc y}^m\rangle^{G_s}/\langle {\mathpzc y}^m\rangle-1|\,.
\end{equation}
In all cases, a value of $j_s=4$ ensures
\begin{equation}
{\rm mean}\{
|\langle {\mathpzc y}^m\rangle^{G_{s}}/\langle {\mathpzc y}^m\rangle^{G_{s-2}}-1|;
m=1,\ldots,m_{\rm max}\}< 1\%.
\end{equation}

\begin{figure}[t]

\includegraphics[clip,width=0.45\textwidth]{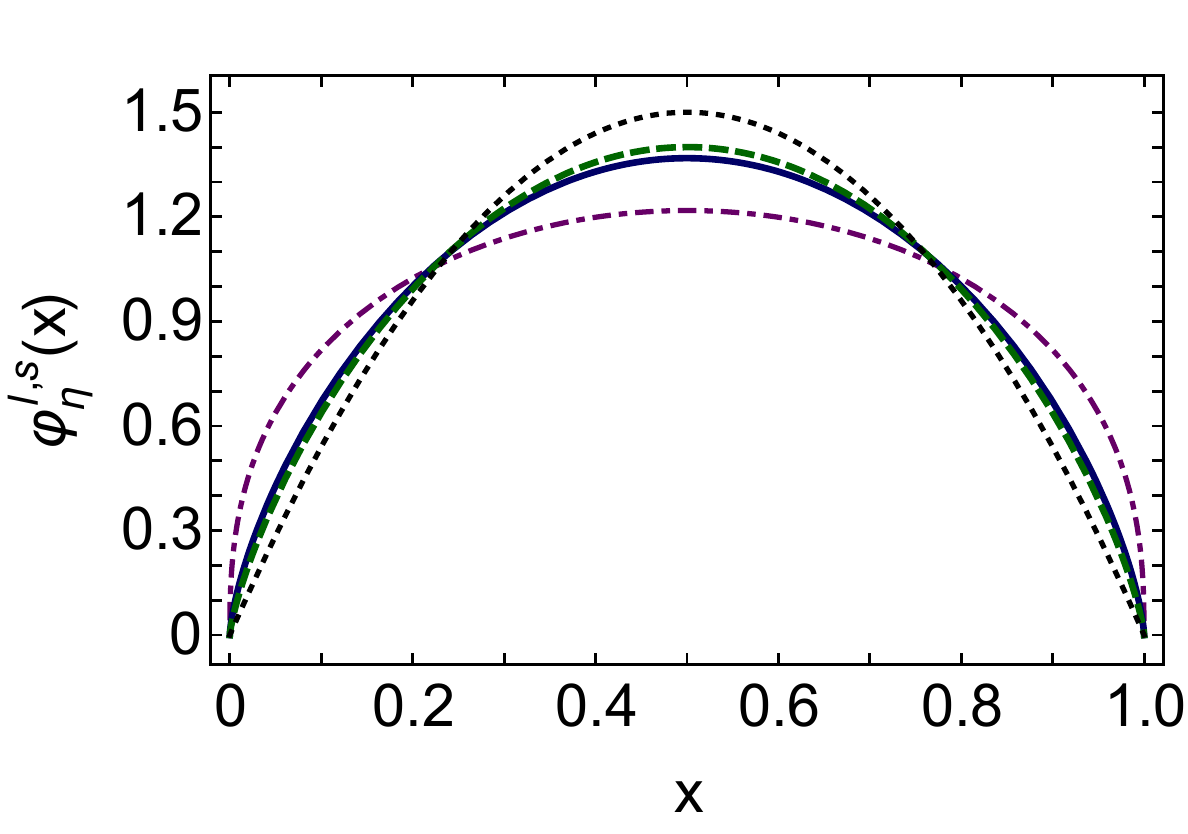}
\includegraphics[clip,width=0.45\textwidth]{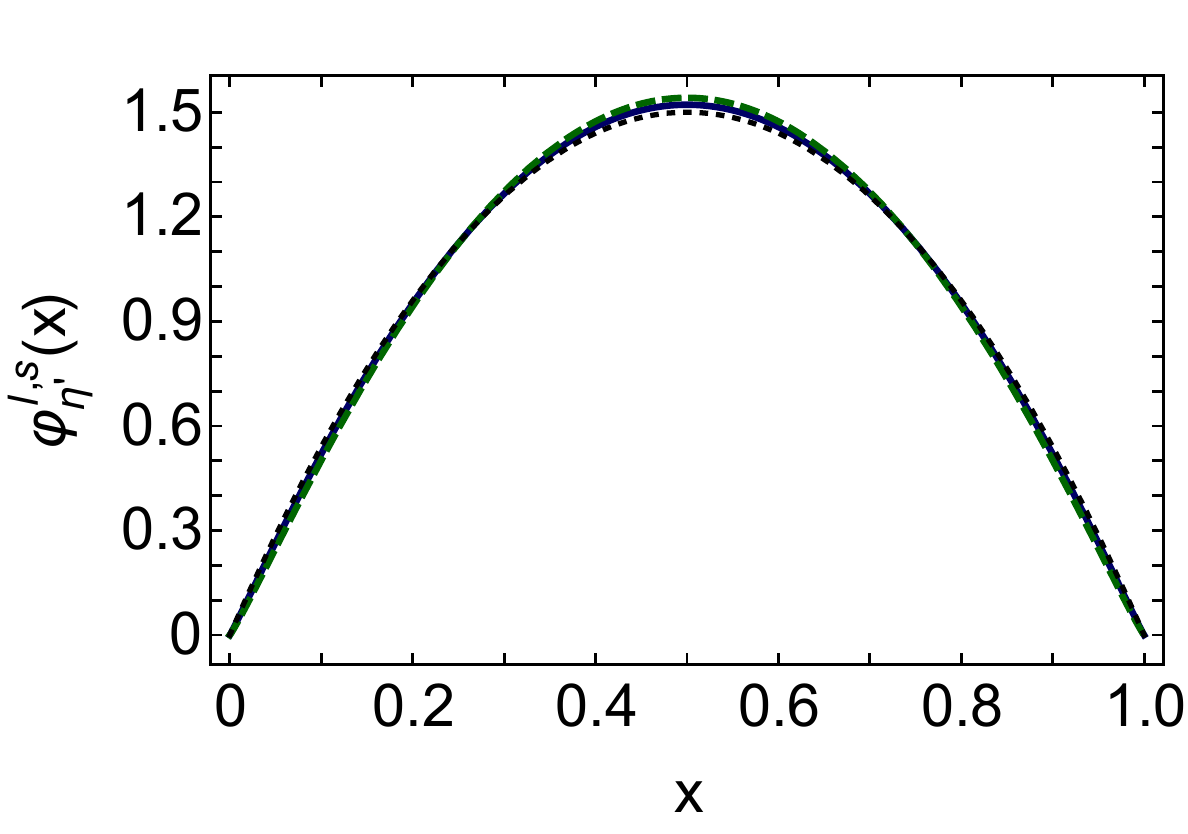}

\caption{\label{figPDAs} Computed light-quark (solid blue curve) and $s$-quark (dashed green curve) component DAs of the $\eta$-meson (upper panel) and $\eta^\prime$-meson (lower panel), determined at $\zeta=2\,{\rm GeV} =:\zeta_2$, listed in Eqs.\,\eqref{mesonPDAs}.
For comparison: upper panel, dot-dashed (purple) curve -- pion's dressed-valence-quark distribution amplitude \cite{Chang:2013nia, Chen:2018rwz}; and both panels, dotted black curve -- asymptotic profile, Eq.\,\eqref{PDAcl}.}
\end{figure}

In each instance, this level of accuracy is achieved with small values of the coefficients $a_{2,4}^\alpha$.  We therefore take an additional step, setting $a_{j\geq 2}^\alpha\equiv 0$ in Eq.\,\eqref{PDAEquation} and reconstructing pointwise approximations to the DAs solely by fitting $\alpha$.  The results obtained in this way are not realistically distinguishable from those determined with the more general procedure.  Hence, in our subsequent analysis we used these simpler forms (depicted in Fig.\,\ref{figPDAs}):
\begin{subequations}
\label{mesonPDAs}
\begin{align}
  \varphi_M^{\,{\mathsf f}}&(x)  = {\mathpzc n}_{\alpha_M^{\,{\mathsf f}}} [ x(1-x) ]^{\alpha_M^{\,{\mathsf f}}},\\
&\begin{array}{c|c|c|c}
\alpha_\eta^l & \alpha_\eta^s & \alpha_{\eta^\prime}^l & \alpha_{\eta^\prime}^s\\\hline
0.70 & 0.77 & 1.05 & 1.10
\end{array}.\qquad
\end{align}
\end{subequations}

It is worth reiterating that the DAs in Fig.\,\ref{figPDAs} are predictions, deriving from Bethe-Salpeter wave functions computed using the same truncation scheme for the continuum bound-state problem that successfully unified the pion's elastic and transition form factors with those of heavier pseudoscalar mesons \cite{Chang:2013nia, Raya:2015gva, Raya:2016yuj, Chen:2018rwz}.  Evidently, within each bound state, the light- and $s$-quark component DAs have very similar profiles; but there are significant differences between the mesons:
the $\eta$-meson DAs are both broader than the asymptotic distribution,
whereas the $\eta^\prime$ distributions are narrower.
Notwithstanding this, all DAs are measurably narrower than that associated with the pion's valence dressed-quark distribution. %
These features are consistent with the analysis in Ref.\,\cite{Ding:2015rkn} and Appendix~\ref{AppendixTopology}.  They reflect structural differences between these systems, owing to an interplay between emergent and explicit mass generation in the Standard Model, which are expressed in their Bethe-Salpeter amplitudes and hence affect the transition form factors on the entire domain of accessible $Q^2$ because QCD evolution is logarithmic \cite{Lepage:1979zb, Efremov:1979qk, Lepage:1980fj}.

To implement such evolution, one simply projects a given DA in Eqs.\,\eqref{mesonPDAs} onto the eigenfunctions of the QCD evolution operator, evolves the associated coefficients, then reconstructs the concave evolved DA.  Using one-loop evolution, this can be accomplished with roughly ten lines of computer code.  An illustration is provided, \emph{e.g}.\ in Ref.\,\cite{Segovia:2013eca}.

\section{\mbox{\bf $\eta$}, \mbox{\bf $\eta^\prime$} Transition Form Factors}
\label{sec4}
\subsection{Triangle diagram}
As outlined in Sec.\,\ref{SecIntroduction}, the $\gamma^\ast\gamma \to \eta,\eta^\prime$ transitions are each described by a single form factor, $G_M(Q^2)$, $M=\eta,\eta^\prime$.  In RL truncation, $G_M(Q^2)$ is obtained from the following expression:
\begin{subequations}
\label{TransitionFFs}
\begin{align}
\nonumber
&\tfrac{e^2}{8\pi^2}\,  \epsilon_{\mu\nu\alpha\beta}\,k_{1\alpha} k_{2\beta}\, G_M(k_1^2,k_1\cdot k_2,k_2^2)\\
& = \tfrac{e^2}{8\pi^2}\,  \epsilon_{\mu\nu\alpha\beta}\,k_{1\alpha} k_{2\beta}\,
[G_M^l(Q^2) + G_M^s(Q^2)] \\
& = {\rm tr}_{\rm D}\int_{d\ell}\bigg[
c_l i\chi_\mu^l(\ell,\ell_1) \Gamma_{M}^l(\ell_1,\ell_2) S_l(\ell_2) i\Gamma_\nu^l(\ell_2,\ell) \nonumber \\
& \qquad+ c_s i\chi_\mu^s(\ell,\ell_1) \Gamma_{M}^s(\ell_1,\ell_2) S_s(\ell_2) i\Gamma_\nu^s(\ell_2,\ell)\bigg]\,,
\end{align}
\end{subequations}
where: the trace is over spinor indices; $\ell_{1}=\ell+k_1$, $\ell_{2} = \ell - k_2$; the kinematic conditions are $k_1^2 = Q^2$, $k_2^2=0$, $(k_1+k_2)^2 = -m_M^2$; and $\Gamma_\nu^{\,{\mathsf f}}$ and $\chi_\mu^{\,{\mathsf f}}$ are, respectively, the flavour-dependent amputated and unamputated dressed-photon-quark vertices.

The photon-quark vertices in Eq.\,\eqref{TransitionFFs} may each be obtained by solving a RL-truncation of the associated inhomogeneous Bethe-Salpeter equation \cite{Maris:1999bh, Roberts:2000aa}; but we adopt a simpler approach, which has hitherto proven effective.  Namely, emulating Refs.\,\cite{Chang:2013nia, Raya:2015gva, Raya:2016yuj}, we use the following \emph{Ans\"atze} for the unamputated vertices, expressed completely in terms of the functions which characterise the dressed-quark propagators, Eq.\,\eqref{SfVS}:
\begin{align}
\nonumber
\chi_\mu^{\,{\mathsf f}}&(k_o,k_i)   =   \gamma_\mu \Delta_{k^2 \sigma_V^{\,{\mathsf f}}} \\
\nonumber
& + [{\mathpzc s}_{\mathsf f} \, \gamma\cdot k_{o}\gamma_{\mu}\gamma\cdot k_{i}   + \bar{\mathpzc s}_{\mathsf f} \gamma\cdot k_{i}\gamma_{\mu}\gamma\cdot k_{o}]
\Delta_{\sigma_V^{\mathsf f}}  \\
\nonumber
& +
[{\mathpzc s}_{\mathsf f}\,(\gamma\cdot k_o \gamma_\mu + \gamma_\mu \gamma\cdot k_i ) \\
& \quad + \bar{\mathpzc s}_{\mathsf f}\,(\gamma\cdot k_i \gamma_\mu + \gamma_\mu \gamma\cdot k_o )
 ]\, i \Delta_{\sigma_S^{\mathsf f}}\,, \label{ChiAnsatz}
%
%
\end{align}
where $\Delta_{F}= [F(k_o^2)-F(k_i^2)]/[k_o^2-k_i^2]$,  $\bar{\mathpzc s}_{\,{\mathsf f}}=1-{\mathpzc s}_{\,{\mathsf f}}$.  Likewise, our \emph{Ans\"atze} for $\Gamma_\nu^{\mathsf f}$, based on Eq.\,(3.84) in Ref.\,\cite{Roberts:1994dr}, are analogues for the amputated vertex.

Owing to the Abelian anomaly \cite{Adler:1969gk, Bell:1969ts, Adler:2004ih}, it is impossible to simultaneously conserve the vector and axial-vector currents associated with the triangle-diagram integral in Eq.\,\eqref{TransitionFFs}.\footnote{The manner by which the chiral-limit version of Eq.\,\eqref{TransitionFFs} provides for a parameter-free realisation of the Abelian anomaly constraint is detailed in Refs.\,\cite{Roberts:1994hh, Maris:1998hc, Maris:2002mz, Holl:2005vu}.}
This has a measurable effect in the neighbourhood of $Q^2=0$ and that is why we have included a momentum redistribution factor in Eq.\,\eqref{ChiAnsatz} \cite{Raya:2015gva}:
\begin{subequations}
\label{s0}
\begin{align}
s_{\mathsf f} & = 1 + s_0^{\mathsf f} \exp(-{\mathpzc E}_{\mathsf f}/M_{\mathsf f}^E)\,,\\
{\mathpzc E}_{\mathsf f} & = [Q^2/4 + (M_{\mathsf f}^E)^2]^{1/2} -M_{\mathsf f}^E\,,
\end{align}
\end{subequations}
where ${\mathpzc E}_{\mathsf f}$ is a Breit-frame kinetic energy
and $M_{\mathsf f}^E=\{p|p^2=M_{\mathsf f}^2(p^2),p>0\}$ is our calculated value of the Euclidean constituent-mass associated with the va\-le\-nce $\mathsf f$-quark in the pseudoscalar meson \cite{Maris:1997tm, Ding:2015rkn},
\begin{equation}\label{MEuclidean}
M_l^E=0.41\,{\rm GeV},\; M_s^E=0.57\,{\rm GeV}.
\end{equation}
%

Up to transverse pieces associated with ${\mathpzc s}_{\mathsf f}$, $\chi_\mu^{\mathsf f}(k_o,k_i)$ and $S_{\mathsf f}(k_o)\Gamma_\mu^{\mathsf f}(k_o,k_i)S_{\mathsf f}(k_i)$ are equivalent.
All differences are power-law suppressed in the ultraviolet; and while Fig.\,\ref{TFFslowQ2} (below) reveals that making them identical might lead to modest improvements in the description of $\gamma^\ast\gamma\to\eta, \eta^\prime$ transitions at infrared momenta, the computational effort would increase substantially.  Since the cost outweighs the gain, we omit this step herein.

Each element in Eq.\,\eqref{TransitionFFs} is now expressed via a PTIR: Sec.\,\ref{Interpolations} and Eqs.\,\eqref{ChiAnsatz}--\eqref{MEuclidean}.  Hence, the computation of $G_M(Q^2)$ reduces to the task of summing a series of terms, all of which involve a single four-momentum integral.  The integrand denominator in every term is a product of $\ell$-quadratic forms, each raised to some power.  Within each term, one uses a Feynman parametrisation in order to combine the denominators into a single quadratic form, raised to the appropriate power.  A suitably chosen change of variables then enables routine evaluation of the four-momentum integration using algebraic methods.  After calculation of the four-momentum integration, evaluation of the individual term is complete after one computes a finite number of simple integrals; namely, the integrations over Feynman parameters and the spectral integral.  The complete result for $G_M(Q^2)$ follows after summing the series.
Following this procedure, one may fix the redistribution factors in Eq.\,\eqref{s0}.   Using
\begin{align}
\Gamma_{M\to\gamma\gamma} & = \frac{1}{4} \pi \alpha_{\rm em}^2 m^3_{M}|G_M(Q^2=0)|^2
\end{align}
and requiring reproduction of the results in Eq.\,\eqref{EqWidths}, then ${\mathpzc s}_0^{l}=1.21$, ${\mathpzc s}_0^{s}=0.48$.

\subsection{Evolution and asymptotic limits}
\label{SecEvolution}
Before displaying our complete results for $G_{\eta,\eta^\prime}(Q^2)$, it is sensible to discuss their asymptotic limits and the implementation and impact of QCD evolution \cite{Lepage:1979zb, Efremov:1979qk, Lepage:1980fj}.  Absent the non-Abelian anomaly, the asymptotic limits of $G_{\eta,\eta^\prime}$ would simply be obtained from Eq.\,\eqref{GPqcl}.  However, as described in Sec.\,\ref{SecBSE}, owing to the anomaly,  the singlet decay constant, $f_0$, and thus $f^{l,s}_{\eta,\eta^\prime}$ in Eq.\,\eqref{GPqcl} depend on $\zeta$.  Writing, for notational convenience,
\begin{equation}
\label{DefFM}
 F_{\eta,\eta^\prime}(Q^2) = \frac{1}{2\pi^2}G_{\eta,\eta^\prime}(Q^2)\,,
\end{equation}
the impact of $f_0\to f_0(\zeta)$ can be exhibited as follows  (again, $\tau^2 = \Lambda_{\rm QCD}^2/Q^2$):
\begin{subequations}
\label{UVzeta}
\begin{align}
Q^2 F_M(Q^2) & \stackrel{\tau \simeq 0}{\approx}
\left. 6 \left[ c_l f_M^l(Q^2) + c_s f_M^s(Q^2)\right] \right|_{\tau=0}\\
& = \left. 2 \left[ c_8 f_M^8 + 2 c_0 f_M^0(Q^2) \right]\right|_{\tau=0}\,,
\end{align}
\end{subequations}
$c_8 = 1/\surd 3$, $c_0=\sqrt{2/3}$, where \cite{Agaev:2014wna}
\begin{equation}
\label{f0zeta}
f_M^0(\zeta^2) = f_M^0(\zeta_2^2) \left( 1 + \frac{2 N_f}{\pi\beta_0} \left[\alpha_S(\zeta^2) - \alpha_S(\zeta_0^2)\right]\right)\,,
\end{equation}
with $\zeta_0$ the scale at which the calculation is normalised.

Using our computed values for the leptonic decay constants at $\zeta_2$, Eqs.\,\eqref{OSmixing}, \eqref{OSfvalues}, and a one-loop running coupling defined consistent with Eq.\,\eqref{defcalG}:
\begin{subequations}
\label{UVvalues}
\begin{align}
\label{UVeta}
Q^2F_\eta(Q^2) & \stackrel{\tau \simeq 0}{\approx} 0.15\,{\rm GeV}\,,\\
\label{UVetaP}
Q^2 F_{\eta^\prime}(Q^2) & \stackrel{\tau \simeq 0}{\approx} 0.30\,{\rm GeV}\,.
\end{align}
\end{subequations}
The evolution of $f_M^0$ in Eq.\,\eqref{f0zeta} reduces the $\eta$-meson limit by 1\% and that of the $\eta^\prime$ by 10\%.  (This does not alter their ordering with respect to the neutral pion, for which the result is $2f_\pi \approx 0.186\,$GeV.)  Notably, our starting scale is fixed: we know the point at which the propagators, vertices and amplitudes are renormalised \emph{i.e}.\ $\zeta_0=\zeta_2$.  This is not the case with estimates based on \emph{Ans\"atze} for the DAs, in which the scale $\zeta_0$ is a model parameter.  Were one to use our computed DAs, but associate them with a scale $\zeta_0=1\,$GeV, then the suppressions would naturally be greater: 1.5\% for the $\eta$ and 15\% for the $\eta^\prime$.  Existing data cannot distinguish between effects on this scale.  On the hand, as we shall see, they are sensitive to the $\zeta=\zeta_0$ values of the decay constants and mixing angles.  Our computed values, Eqs.\,\eqref{MixingFit}, \eqref{OSfvalues}, control the $Q^2$-dependence of the results for $F_{\eta,\eta^\prime}$ because they are encoded in the Bethe-Salpeter wave functions for these bound states.

The evolution of a flavour nonsinglet DA with the resolving scale $\zeta$ is explained in Refs.\,\cite{Lepage:1979zb, Efremov:1979qk, Lepage:1980fj}.  It is logarithmic; and whilst Poincar\'e covariant computations using a renormalisation-group-improved RL truncation produce the same matrix-element power-laws as perturbative QCD, they fail to reproduce the full anomalous dimensions \cite{Lepage:1980fj}.  Typically \cite{Maris:1998hc, Chang:2013pq, Chang:2013nia}, the RL approximation to a form factor, such as that defined by Eq.\,\eqref{TransitionFFs}, underestimates the rate of its logarithmic flow with the active momentum scale because the approximation omits gluon-splitting diagrams.

As explained elsewhere \cite{Raya:2015gva}, this defect of RL truncation can be corrected by recognising that, owing to Eq.\,\eqref{MesonPDA}, a given meson's Poincar\'e covariant wave function must evolve with $\zeta$ in the same way as its DA.
Such evolution enables the dressed-quark and -antiquark degrees-of-freedom, in terms of which the wave function is expressed at a given scale $\zeta^2=Q^2$, to split into less well-dressed partons via the addition of gluons and sea quarks in the manner prescribed by QCD dynamics.  These effects are incorporated naturally in bound-state problems when the complete quark-antiquark scattering kernel is used; but aspects are lost when that kernel is truncated, and so it is with the truncation used herein.

Similar to the decay constants, the non-Abelian anomaly complicates evolution of the DAs and Bethe-Salpeter wave functions of the $\eta$- and $\eta^\prime$-mesons.  This is most readily explained by shifting to the octet-singlet basis at $\zeta_2$:
\begin{subequations}\label{PDAOSbasis}
\begin{align}
f_M^8 \varphi_M^8 &= c_8 f_M^l \varphi_M^l - c_0 f_M^s \varphi_M^s\,,\\
f_M^0 \varphi_M^0 &= c_0 f_M^l \varphi_M^l + c_8 f_M^s \varphi_M^s\,.
 \end{align}
\end{subequations}
Defined in this way, $\varphi_M^8(x)$ evolves without mixing at leading-order.  However, $\varphi_M^0(x)$ mixes with a two-gluon DA, $\varphi_M^g$, under leading-order evolution \cite{Agaev:2014wna}.  To implement the effect, one would need either to compute $\varphi_M^g$ or employ a reliable model.  No calculations are currently available and few constraints exist that can be used to aid in developing a good \emph{Ansatz}.  Hence, we set $\varphi_{\eta,\eta^\prime}^g\equiv 0$ and thereby suppress mixing.  Whilst this may seem a drastic step, considering the impact of analogous effects on the decay constants and the analysis in Ref.\,\cite{Agaev:2014wna}, we expect it to have little impact on $F_\eta$  and to introduce an uncertainty of only $\pm 10$\% in $F_{\eta^\prime}$ on the empirically accessible domain.
This uncertainty also absorbs any contribution from a $c\bar c$ component in the $\eta$, $\eta^\prime$-meson Bethe-Salpeter wave functions.  In any event, we expect this type of intrinsic charm contribution to be small, owing to the quark-line hairpin structure of the anomaly kernel, Fig.\,\ref{FigAnomaly}, which suppresses such mixing \cite{Bhagwat:2007ha}, and the absence of any need for an intrinsic light-quark component in describing the $\gamma^\ast\gamma \to \eta_c$ transition \cite{Raya:2016yuj}.
Since $m_b \gg m_c\gg m_s$, any contribution to $F_{\eta,\eta^\prime}$ from a $b\bar b$ component is implausible.

Following these observations, we implement evolution of the Bethe-Salpeter amplitudes (and transition form factors) as follows.  (\emph{i})  Write
\begin{equation}
\chi_M^8  = c_8 \chi_M^l - c_0 \chi_M^s\,,\;
\chi_M^0  = c_0 \chi_M^l + c_8 \chi_M^s\,,
\end{equation}
and hence, equally,
\begin{subequations}
\begin{align}
F_M^8(Q^2)  & = c_8 F_M^l(Q^2) - c_0 F_M^s(Q^2)\,,\;\\
F_M^0(Q^2)  & = c_0 F_M^l(Q^2) + c_8 F_M^s(Q^2)\,;
\end{align}
\end{subequations}
(\emph{ii}) on $Q^2> \zeta_2^2$,
\begin{subequations}
\begin{align}
F_M^8(Q^2) & \rightarrow  F_M^8(Q^2) [\kappa_M^{8}(Q^2)/\kappa_M^{8}(\zeta_2^2)]\,,\\
F_M^0(Q^2) & \rightarrow  F_M^0(Q^2) [f_M^0(Q^2)\kappa_M^{0}(Q^2)/ f_M^0(\zeta_2^2)\kappa_M^{0}(\zeta_2^2)]\,,
\end{align}
\end{subequations}
where
\begin{equation}
\kappa_M^{8,0}(Q^2) = \tfrac{1}{2}\int_0^1 dx\, \varphi_M^{8,0}(x;Q^2) [ 1/x + (2x -1)^2 ] \,;
\end{equation}
and (\emph{iii}) rebuild the transition form factors in the quark flavour basis, \emph{viz}.\
\begin{subequations}
\begin{align}
F_M^l(Q^2)  &= \phantom{-} c_8 F_M^8(Q^2) + c_0 F_M^0(Q^2)\,,\\
F_M^s(Q^2) &= - c_0 F_M^8(Q^2) + c_8 F_M^0(Q^2)\,,
\end{align}
\end{subequations}
and therefrom, using Eqs.\,\eqref{TransitionFFs}, the final results for $F_M(Q^2)$.  In this way, we generalise the procedure introduced and explained in Refs.\,\cite{Raya:2015gva, Raya:2016yuj}.

\subsection{Calculated transition form factors}
\subsubsection{Low $Q^2$}
\begin{figure}[t]

\includegraphics[clip,width=0.45\textwidth]{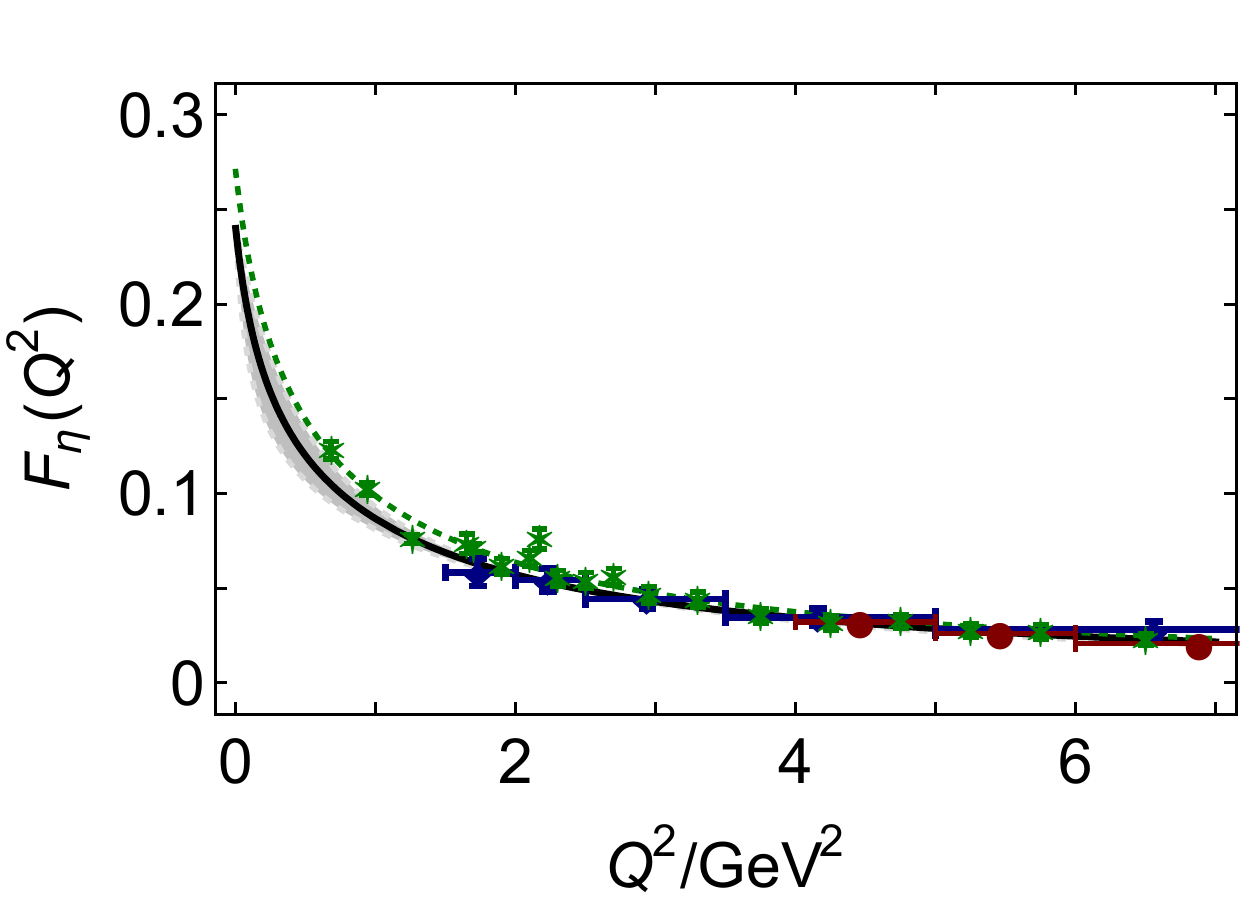}
\includegraphics[clip,width=0.45\textwidth]{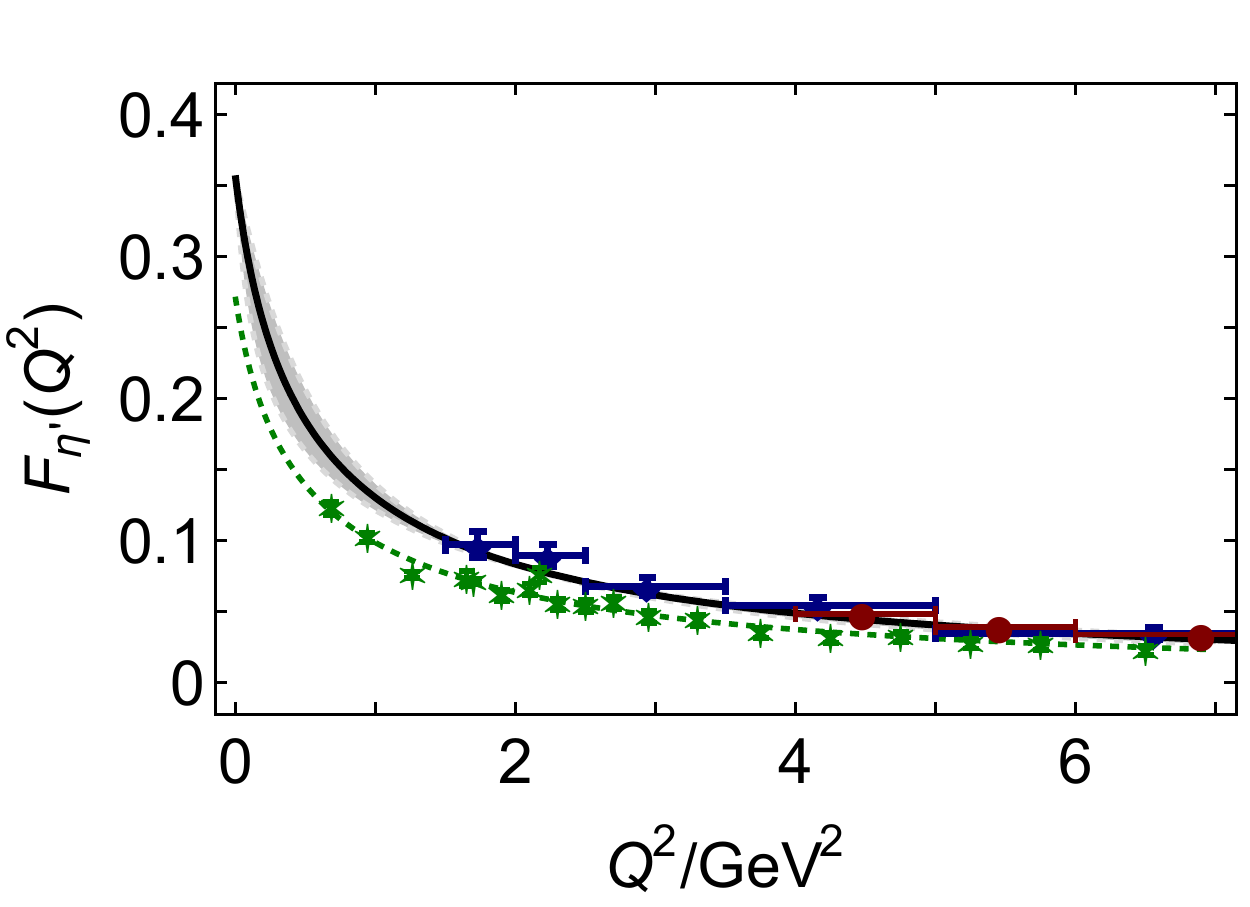}

\caption{\label{TFFslowQ2} $\gamma^\ast \gamma \to \eta,\eta^\prime$ transition form factors, normalised according to Eq.\,\eqref{DefFM}: upper panel, $\eta$; lower panel, $\eta^\prime$.
Solid (black) curves -- our predictions, with the bracketing (grey) bands indicating an estimate of uncertainty associated with the photon-quark vertex \emph{Ans\"atze} in Eq.\,\eqref{TransitionFFs}.
Dotted (green) curve -- for comparison, $\pi^0$ result computed in Ref.\,\cite{Raya:2015gva}; and associated experimental data -- (green asterisks) from ``CELLO'' \cite{Behrend:1990sr} and ``CLEO'' \cite{Gronberg:1997fj}.
The $\eta, \eta^\prime$ data are: diamonds (blue) CLEO \cite{Gronberg:1997fj}; circles (red) ``BaBar'' \cite{BABAR:2011ad}.
}
\end{figure}

%
We have computed $F_{\eta,\eta^\prime}(Q^2)$ using the inputs and method described above; and in Fig.\,\ref{TFFslowQ2} we depict their behaviour on a low-$Q^2$ domain: our predictions are indicated by the solid (black) curves in each panel.\footnote{%
The PTIRs detailed in Appendix~\ref{AppendixA} enable a direct evaluation of the integral in Eq.\,\eqref{TransitionFFs} on $P^2>-(0.85\,{\rm GeV})^2$.  For the $\eta^\prime$, we therefore evaluate this integral on $P^2/{\rm GeV}^2 \in (-0.85^2,-0.75^2)$ and extrapolate to $P^2=-(0.96\,{\rm GeV})^2$ using $[n,n]$ Pad\'e approximants, $n=1,2,3,4$, with the difference between the extrapolated values being used to estimate the error in the procedure.  That error is small, lying within the line-width of all $\eta^\prime$ curves.}

In these calculations we used \emph{Ans\"atze} for $\chi_\mu^{\mathsf f}$, $\Gamma_\nu^{\mathsf f}$ in Eq.\,\eqref{TransitionFFs} instead of solutions of the relevant inhomogeneous Bethe-Salpeter equations.  Whilst they are efficacious, with longitudinal parts constrained completely by Ward-Green-Takahashi identities, they are not perfect: as noted above, there are model uncertainties in the transverse pieces which may affect the $Q^2$-dependence of the results on $Q^2\lesssim m_V^2$, where $m_V$ is the appropriate vector meson mass ($m_\rho \approx 2 M^E_l$, $m_\phi \approx 2 M^E_s$).
Notably, however, such uncertainties had neither a visible impact on the $\gamma^\ast \gamma \to \pi^0, \eta_c, \eta_b$ transition form factors \cite{Raya:2015gva, Raya:2016yuj} nor on the charged-pion and -kaon elastic form factors \cite{Chang:2013nia, Gao:2017mmp}.  Relative to these systems, a difference herein is the non-Abelian anomaly, which affects $\eta, \eta^\prime$ structure and conceivably, therefore, generates corrections to Eq.\,\eqref{TransitionFFs} at infrared momenta. This will be explored elsewhere.
%
Here, we simply estimate the sensitivity of our results to neglected infrared effects by supposing that any such uncertainty is maximal on $Q^2 \simeq (M^E_l+M^E_s)^2/4$ (as much as 30\% in total) and vanishes smoothly either side of this domain because: (\emph{i}) the $\eta,\eta^\prime \to \gamma\gamma$ widths constrain $F_{\eta,\eta^\prime}(Q^2=0)$; and (\emph{ii}) our form factor predictions match data on $Q^2\gtrsim 2\,$GeV$^2$.  This procedure produces the (grey) bands in the panels of Fig.\,\ref{TFFslowQ2}.

Globally, the sensitivity to potential infrared corrections is negligible; but it is apparent in our estimates of the interaction radii:
\begin{align}
\label{DefineRadii}
r_M^2 &:= \left. \frac{-6}{F_M(0)} \frac{d}{dQ^2} F_M(Q^2)\right|_{Q^2=0}\,,\\
r_{\eta} &= 0.83^{+0.40}_{-0.22}\,{\rm fm},\;
r_{\eta^\prime} = 0.73^{+0.34}_{-0.19}\,{\rm fm}.
\end{align}
Empirically, extracted from measurements of the Dalitz decays $\eta,\eta^\prime \to \gamma e^+ e^-$:
$r_{\eta} = 0.67(3)\,$fm \cite{Aguar-Bartolome:2013vpw},
$r_\eta^\prime = 0.61(3)\,$fm \cite{Ablikim:2015wnx}.  Our calculated value for the ratio
\begin{equation}
r_{\eta}/r_{\eta^\prime} = 1.14(1)\,,
\end{equation}
which matches experiment: $1.10(7)$, has little uncertainty because any change in the computed value of one radius is compensated by that in the other.

\begin{figure}[t]

\includegraphics[clip,width=0.45\textwidth]{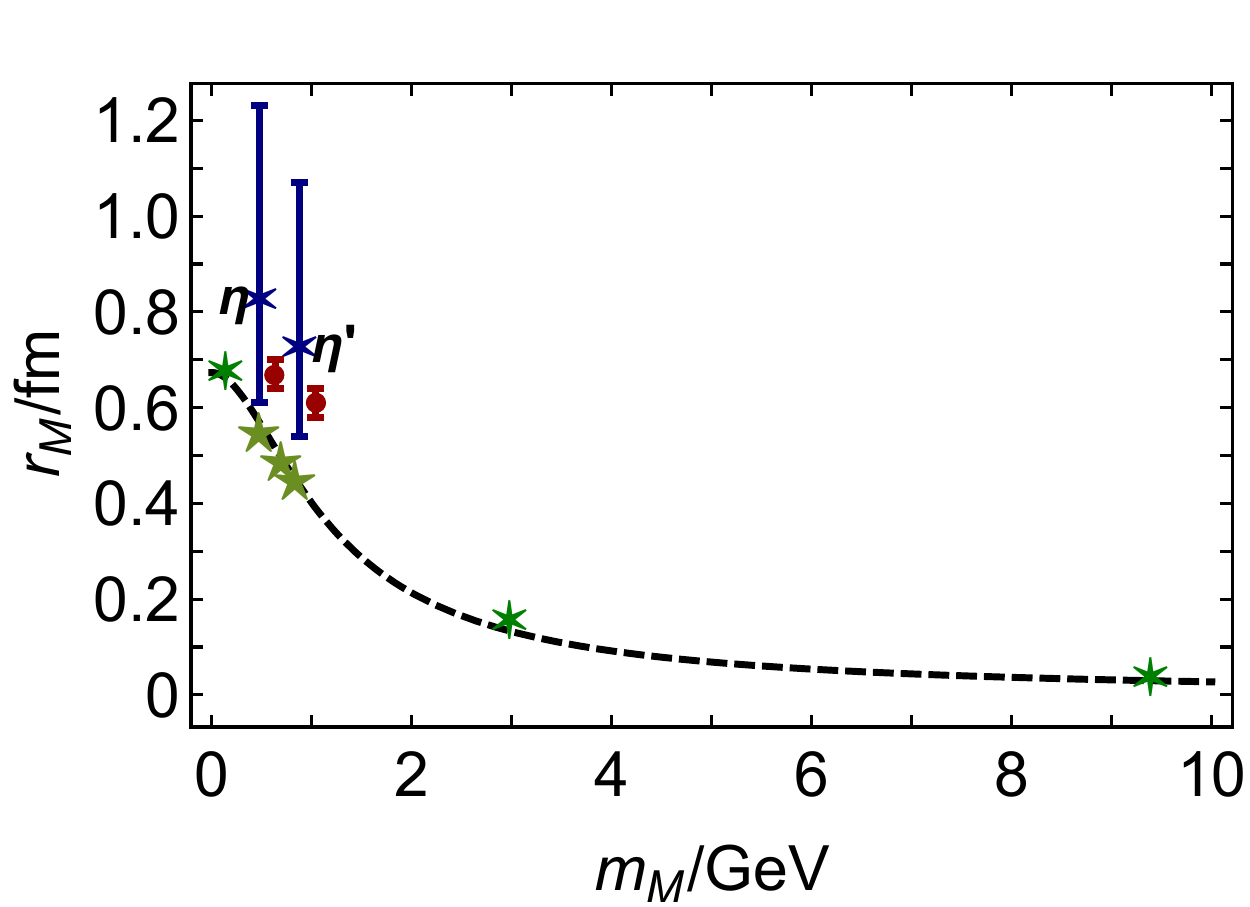}

\caption{\label{FigRadii} Interaction radii of neutral pseudoscalar mesons, Eq.\,\eqref{DefineRadii}, plotted versus meson mass.
For clarity, the $\eta,\eta^\prime$ values are labelled and offset slightly from their true masses: our results (blue asterisks) are compared with empirical estimates (red circles), drawn from Refs.\,\cite{Aguar-Bartolome:2013vpw, Ablikim:2015wnx}.
The $\pi^0$, $\eta_c$, $\eta_b$ values (green asterisks) are taken from Refs.\cite{Raya:2015gva, Raya:2016yuj}.
Five-point stars (olive): computed charge radii of pion-like mesons with masses $m_{0^-}/{\rm GeV}=0.47, 0.69, 0.83$ \cite{Chen:2018rwz}.
Notably, all radii are well described by Eq.\,\eqref{FitRadii} (dashed black curve) except those of $\eta,\eta^\prime$.}
\end{figure}

The calculated $\eta, \eta^\prime$ interaction radii are plotted in Fig.\,\ref{FigRadii}.  We also include another set, \emph{viz}.\ those of different neutral pseudoscalar mesons for which the transition form factors have been computed ($\pi^0$, $\eta_c$, $\eta_b$) \cite{Raya:2015gva, Raya:2016yuj} and the electric-charge radii of pion-like mesons with masses $m_{0^-}/{\rm GeV}=0.47, 0.69, 0.83$ \cite{Chen:2018rwz}: where comparisons are possible, the charge-radii agree with those computed using lattice-QCD \cite{Chambers:2017tuf, Koponen:2017fvm}.  The dashed curve in Fig.\,\ref{FigRadii} is the following interpolation of these additional results:
\begin{equation}
\label{FitRadii}
r_M(m_M) = \frac{r_0}{1+(m_M/{\mathpzc m}) \ln[1+m_M/{\mathpzc m}]}\,,
\end{equation}
where $r_0=0.67\,$fm, ${\mathpzc m}=1.01\,{\rm GeV}\approx m_\phi$.  The kaon point \cite{Gao:2017mmp}: $({\rm mass}=0.49\,{\rm GeV},\mbox{{\rm charge-radius}}=0.58\,{\rm fm})$,  also sits on this curve.  Evidently, for systems not affected by the non-Abelian anomaly, a standard pattern of damping with Higgs-generated current-quark mass is established once $m_M$ exceeds that of the (fictitious) $s \bar s$ bound-state \cite{Ding:2015rkn}.
On the other hand, the $\eta, \eta^\prime$ interaction radii do not fit this pattern: they are larger by 24\% and 48\%.  This effect is greater in the $\eta^\prime$, which is predominantly a $U(3)$ flavour-singlet state and, hence, that system most affected by the non-Abelian anomaly.  (See Appendix~\ref{AppendixTopology}.)

\begin{figure}[t]

\includegraphics[clip,width=0.45\textwidth]{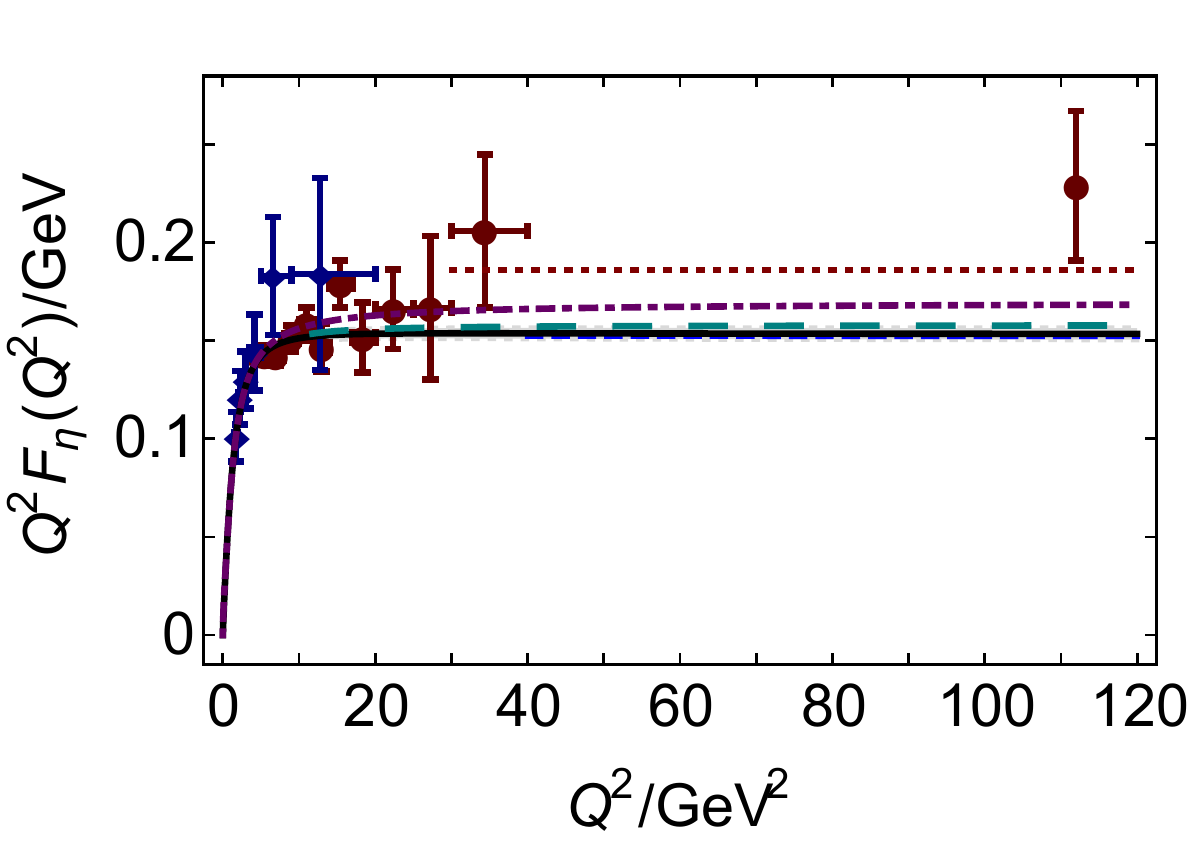}
\includegraphics[clip,width=0.45\textwidth]{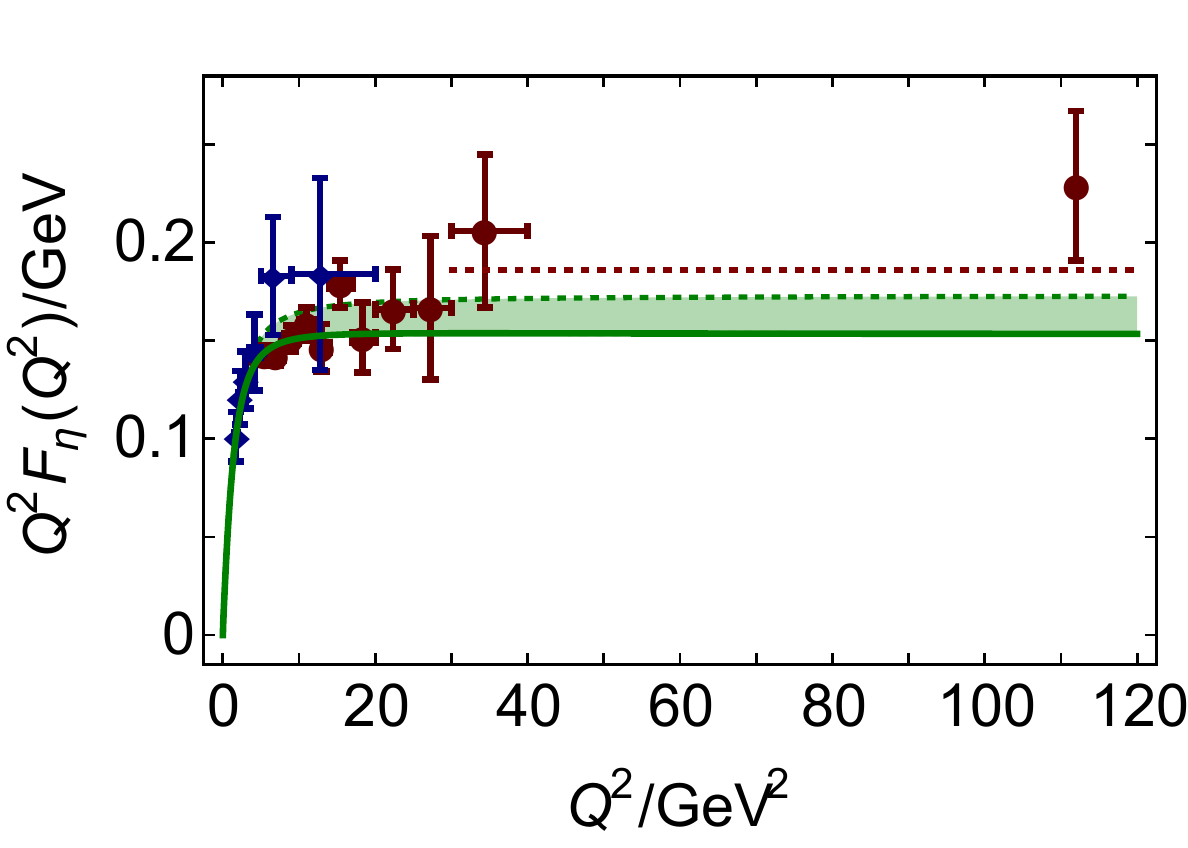}

\caption{\label{FigFM1} $\gamma^\ast \gamma \to \eta$ transition form factor, normalised according to Eq.\,\eqref{DefFM}.
\emph{Upper panel}.  Curves: solid (black), our prediction with complete evolution, described in Sec.\,\ref{SecEvolution}; dot-dashed (purple), results without evolution; long-dashed (cyan), 1-loop evolution only; and dashed (blue), asymptotic limit from Eqs.\,\eqref{f0zeta}, \eqref{UVvalues}.
The grey bands bracketing our full prediction indicate the uncertainty owing to omission of $\varphi_M^0$-$\varphi_M^g$ mixing: in this case, it is negligible.
\emph{Lower panel}.  The shaded (green) band indicates an uncertainty in our prediction for $F_{\eta}$ owing to variations in the value of $f_\eta^s$ (see text).
In both panels, the dotted (red) curve is the $\pi^0$ asymptotic limit, $2f_\pi$;
and the data are: diamonds (blue) CLEO \cite{Gronberg:1997fj}; circles (red) BaBar \cite{BABAR:2011ad, Aubert:2006cy}, where the timelike datum from the latter is plotted at $Q^2=-q^2$.
}
\end{figure}

\subsubsection{Large $Q^2$}
\begin{figure}[t]

\includegraphics[clip,width=0.45\textwidth]{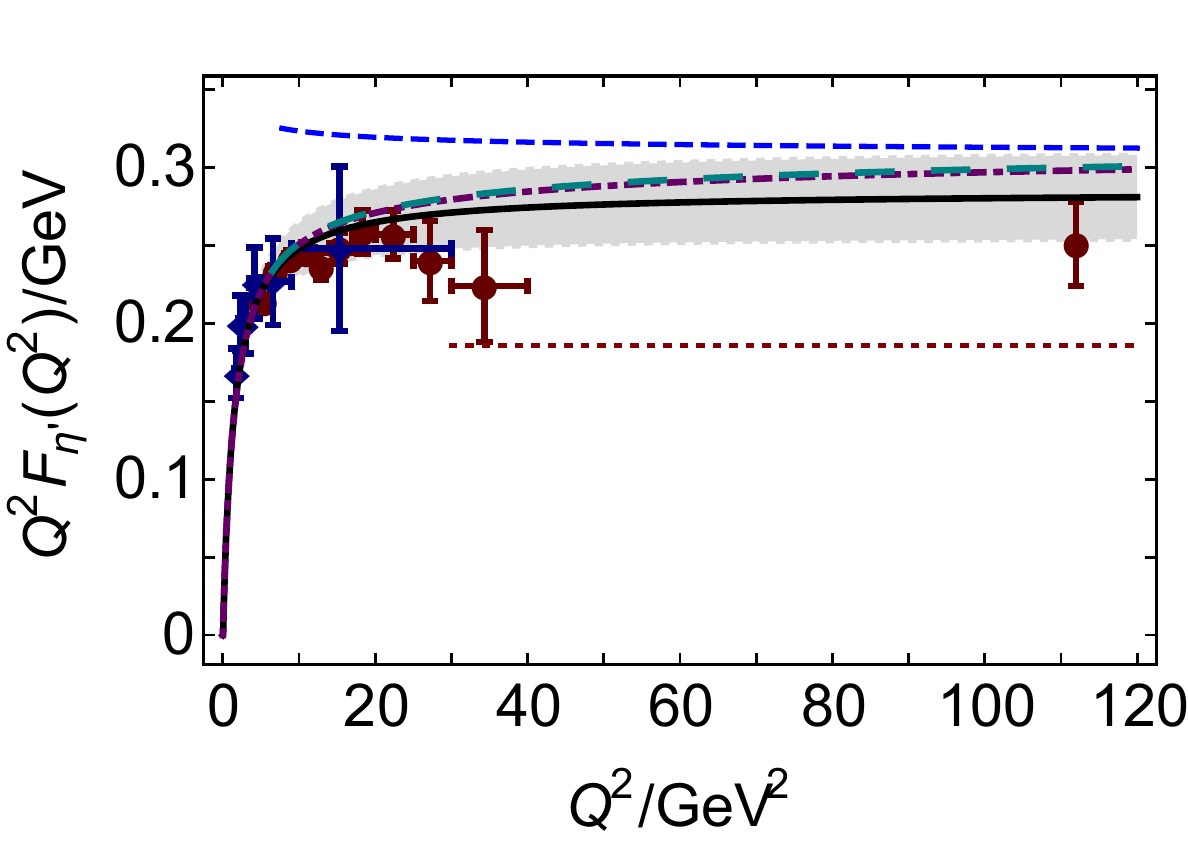}
\includegraphics[clip,width=0.45\textwidth]{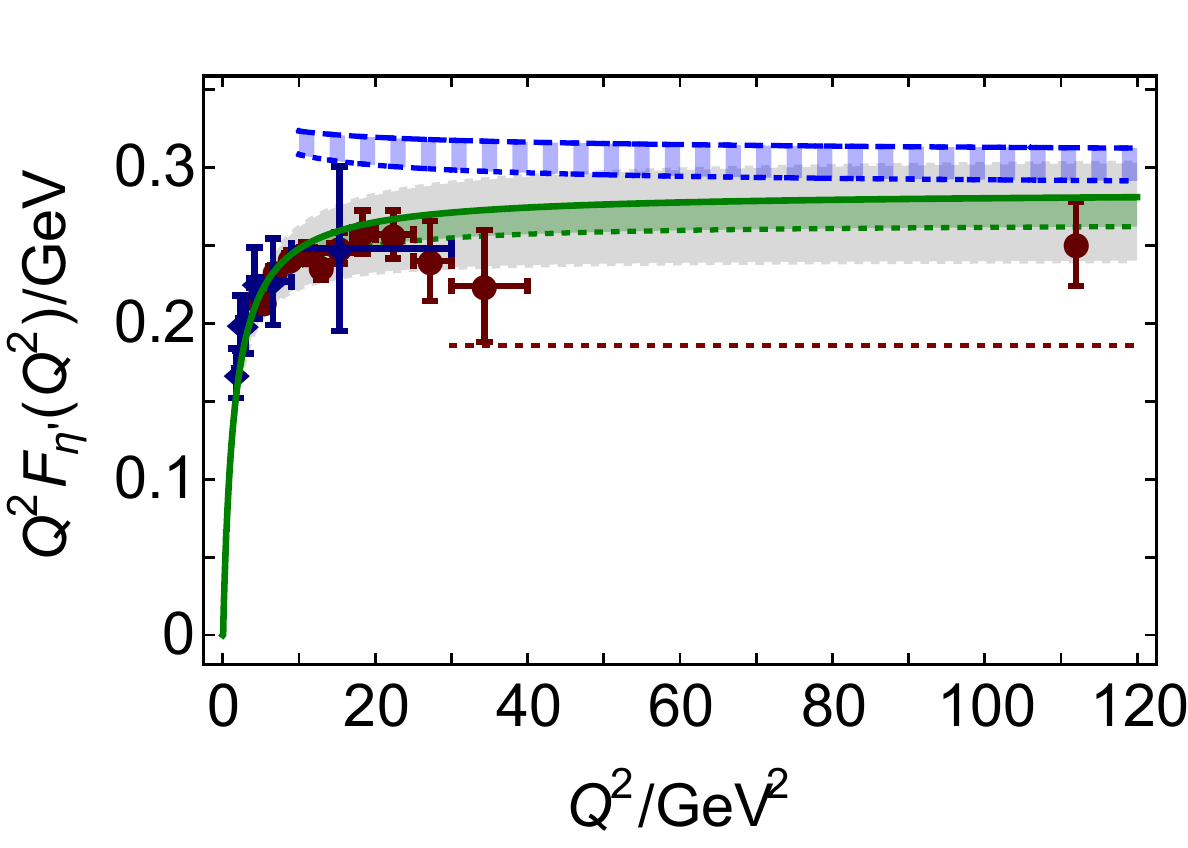}

\caption{\label{FigFM2} $\gamma^\ast \gamma \to \eta^\prime$ transition form factors, normalised according to Eq.\,\eqref{DefFM}.
\emph{Upper panel},
Curves: solid (black), our prediction with complete evolution, described in Sec.\,\ref{SecEvolution}; dot-dashed (purple), results without evolution; long-dashed (cyan), 1-loop evolution only; and dashed (blue), asymptotic limit from Eqs.\,\eqref{f0zeta}, \eqref{UVvalues}.
The grey bands bracketing our full prediction indicate an uncertainty owing to omission of $\varphi_M^0$-$\varphi_M^g$ mixing.
\emph{Lower panel}.
The shaded (green) band indicates an uncertainty in our prediction for $F_{\eta^\prime}$ owing to variations in the value of $f_\eta^s$ (see text).  The broader, shaded (grey) band combines this with the uncertainty owing to omission of  $\varphi_M^0$-$\varphi_M^g$ mixing.
The (blue) banded-shading indicates the impact of uncertainty in $f_\eta^s$ on the asymptotic behaviour of $F_{\eta^\prime}$, Eq.\,\eqref{UVetaP}.
%
In both panels, the dotted (red) curve is the $\pi^0$ asymptotic limit, $2f_\pi$;
and the data are: diamonds (blue) ``CLEO'' \cite{Gronberg:1997fj}; circles (red) ``BaBar'' \cite{BABAR:2011ad, Aubert:2006cy}, where the timelike datum from the latter is plotted at $Q^2=-q^2$.
}
\end{figure}

Our predictions for the large-$Q^2$ behaviour of the transition form factors are depicted in Figs.\,\ref{FigFM1}, \ref{FigFM2}: with the normalisation in Eq.\,\eqref{DefFM}, the asymptotic value of the $\pi^0$ form factor is $2 f_\pi = 0.186\,$GeV, drawn as the dotted (red) curve in all panels.

Consider first $F_{\eta}(Q^2)$ in Fig.\,\ref{FigFM1}.
There are marked similarities with the $\pi^0$ transition form factor (see Ref.\,\cite{Raya:2015gva}, Fig.\,2.).  Namely, the asymptotic limit, Eq.\,\eqref{UVeta}, is only slightly exceeded on $Q^2 \gtrsim 13\,$GeV$^2$; and, including necessary evolution of the meson wave function, that limit is approached uniformly from above with increasing momentum transfer.

Further, our full $F_{\eta}(Q^2)$ result (solid, black curve) agrees well with existing data \cite{Gronberg:1997fj, BABAR:2011ad, Aubert:2006cy}.  Looking at details, it might appear that there is a mismatch between our curve and the largest-$Q^2$ CLEO and BaBar results.  Pursuing this, a review of the results in Table~\ref{TableB} and Eqs.\,\eqref{MixingFit} may suggest that our prediction for the magnitude of $f_\eta^s$ is $\sim 15$\% too high.  To explore the impact of such an overestimate, we changed $f_\eta^s \to 0.85\,f_\eta^s$ and recomputed both the width in Eq.\,\eqref{eqWidthseta} and asymptotic limit in Eq.\,\eqref{UVeta}, with the results $0.42 \to 0.46\,$keV and $0.15\to 0.17\,$GeV, respectively.  Such a 10\% increase in $\Gamma_{\eta\to\gamma\gamma}$ would bring our result into better agreement with experiment.  Hence, the increase in Eq.\,\eqref{UVeta} might also be realistic.  It could be achieved by fine-tuning the parameters that specify ${\mathpzc K}_A$ in Eq.\,\eqref{EqGenK}, listed in Table~\ref{TableA}.  Instead of doing that, however, in the lower panel of Fig.\,\ref{FigFM1} we choose to place an uncertainty on our prediction for $F_{\eta}(Q^2)$; namely, the shaded (green) band.  Evidently, this does not materially affect the comparison with data; so we remain confident of our prediction.
A plausible conclusion is that the largest-$Q^2$ BaBar datum \cite{Aubert:2006cy} is too large by $\sim 50$\%.  This would resolve the mismatch with our prediction and solve the puzzle of its near equality in magnitude with the analogous $\eta^\prime$ datum, which is otherwise difficult to explain.

Turning now to Fig.\,\ref{FigFM2}, upper panel, the behaviour of $F_{\eta^\prime}(Q^2)$ also matches expectations based upon studies of $\gamma^\ast \gamma \to \pi^0, \eta_c, \eta_b$, \emph{e.g}.\ as with $\gamma^\ast \gamma \to \eta_c, \eta_b$, here the asymptotic limit is approached uniformly from below.
The new feature is the impact of the scale dependence of $f_{\eta^\prime}^0$.  It generates a suppression of the transition form factor, which serves to improve the agreement between our result and available experimental data.  At asymptotically large momentum transfers, \emph{i.e}.\ on $\tau \simeq 0$, our full result (solid black curve) meets the asymptotic trajectory (dashed blue curve).

Any overestimate of the size of $f_\eta^s$ also has an impact on $F_{\eta^\prime}$ through its effect on $f_0$, $\theta_0$ in Eq.\,\eqref{OSmixing}.  This is illustrated in the lower panel of Fig.\,\ref{FigFM2}.
For $f_\eta^s \to 0.85 f_\eta^s$, the asymptotic limit in Eq.\,\eqref{UVetaP} is reduced: $0.30\to 0.28\,$GeV, indicated by the (blue) banded-shading extending below the dashed (blue) curve.
Likewise, our prediction for $F_{\eta^\prime}(Q^2)$ is suppressed, as shown by the (green) shading extending below the solid (green) curve.
Our estimate for the combined effect of $\leq 15$\% uncertainty in $f_\eta^s$ and omitting $\varphi_M^0$-$\varphi_M^g$ mixing is represented by the broad grey band.
Within errors, there is agreement between our prediction and all data on the $\gamma^\ast \gamma \to \eta^\prime$ transition.

\section{Epilogue}
\label{SecEpilogue}
Conscious of their importance in validating QCD hard scattering formulae; a need to unify their analysis with the transition form factors of other neutral pseudoscalar mesons and thereby identify remaining challenges to achieving a sound global understanding; and the possibility of much more data from new-generation $e^+ e^-$ colliders \cite{Nature:2014China, Kou:2018nap, Redmer:2018gah};
we employed a continuum approach to the hadron bound-state problem to calculate the $\gamma^\ast \gamma \to \eta, \eta^\prime$ transition form factors.

Our starting point was the Bethe-Salpeter kernel used successfully to explain the $\gamma^\ast\gamma\to \pi^0, \eta_c, \eta_b$ transitions.  (The same kernel was used with equal success in many other applications, \emph{e.g}.\ charged-pion and -kaon elastic form factors \cite{Chang:2013nia, Gao:2017mmp} and nucleon observables \cite{Wang:2018kto}.)  We augmented this with a four-parameter model for the contribution to this kernel deriving from the non-Abelian anomaly, an improvement necessary for any computation of $\eta, \eta^\prime$ properties: the parameters were fixed by requiring that the solutions of the coupled-channels bound-state problems reproduce the empirical $\eta, \eta^\prime$ masses and the four phenomenologically-determined values of the light- and strange-quark $\eta$, $\eta^\prime$ decay constants.

With the bound-state kernel thus defined, we delivered predictions for:
the $\eta,\eta^\prime \to \gamma\gamma$ decay widths -- Sec.\,\ref{MassesWidths};
the four amplitudes that characterise the light-front longitudinal momentum distributions of the light- and strange-quarks within the $\eta, \eta^\prime$ -- Sec.\,\ref{Distributions};
the $\gamma^\ast \gamma \to \eta, \eta^\prime$ transition form factors,
the associated electromagnetic interaction radii and, at the other extreme, their large-$Q^2$ limits -- Sec.\,\ref{sec4}.
Where available, our results compare favourably with existing data.  Important to this at large-$Q^2$ is a sound understanding and implementation of QCD evolution, which has a visible impact on the $\eta^\prime$.
Furthermore, our analysis provides some novel insights into the properties of $\eta, \eta^\prime$ mesons and associated observable manifestations of the non-Abelian anomaly.


This completes a unified description of a large array of pseudoscalar meson properties, ranging from low-energy $\pi\pi$ scattering \cite{Bicudo:2001jq} to the large-$Q^2$ behaviour of the transition form factors of heavy-heavy systems \cite{Raya:2016yuj}, and visiting many stations in between, \emph{e.g}.\ Refs.\,\cite{Horn:2016rip, Gao:2017mmp}.  The related body of analysis delivers an understanding of the distribution of valence-quarks within mesons that smoothly joins Goldstone modes, constituted from the lightest-quarks in Nature, with systems that are markedly affected by the non-Abelian anomaly and hence topological features of QCD, and also mesons containing the heaviest valence-quarks that can today be studied experimentally.  The positive comparison with data in all sectors confirms that the dressed-valence-quark DAs of light-quark mesons are dilated with respect to the asymptotic profile, $\varphi_{\infty}(x)=6 x(1-x)$; those of systems affected by the anomaly may usefully be approximated by $\varphi_{\infty}$; and those for heavy-heavy systems are compressed, becoming narrower as the current-mass of the valence-quarks increases at any given resolving scale.

This particular study can nevertheless be improved.   Most immediately, by developing better \emph{Ans\"atze} for the photon-quark vertices used in computing the transition form factors, \emph{e.g}.\ by implementing features of solutions of the associated inhomogeneous Bethe-Salpeter equations or by using such solutions directly, possibly after building their perturbation theory integral representations; and also by analysing the impact of corrections induced by the non-Abelian anomaly to our approximation for the $\gamma^\ast \gamma \to \eta, \eta^\prime$ transition current.  These steps would enable reliable predictions to be made for the timelike behaviour of those transition form factors that are accessible via the related Dalitz decays, for which contemporary data exist \cite{Aguar-Bartolome:2013vpw, Ablikim:2015wnx, TheNA62:2016fhr}.

The scope of the analysis herein could also be extended to include the doubly off-shell $\gamma^\ast(k_1) \gamma^\ast(k_2) \to M$ transition form factors, where $M$ is any neutral pseudoscalar meson.  This process, too, is described by only one form factor, $F_{\gamma^\ast M}(k_1^2,k_2^2)$.
However, whereas vector meson dominance (VMD) models and QCD-connected analyses both describe the same large-$Q^2$ behaviour for the singly off-shell form factor, \emph{viz}.\ $F_{\gamma^\ast M}(Q^2,0) \sim 1/Q^2 \sim F_M(Q^2)$, albeit with different normalisations, the large-$Q^2$ predictions of VMD \cite{Landsberg:1986fd} and QCD \cite{Chase:1979ck, Brodsky:1989pv, Anikin:1999cx} for the doubly off-shell form factor are distinctively different:
\begin{subequations}
\begin{align}
& F_{\gamma^\ast M}^{\rm VMD}(k_1^2,k_2^2) \sim 1/(k_1^2 k_2^2) \\
\emph{cf}.\ &
F_{\gamma^\ast M}^{\rm QCD}(k_1^2,k_2^2) \sim 1/(k_1^2+k_2^2)\,.
\end{align}
\end{subequations}
The first data that can distinguish between these predictions now exist for $\gamma^\ast(k_1) \gamma^\ast(k_2) \to \eta^\prime$  \cite{BaBar:2018zpn}.  They favour the QCD result, with behaviour qualitatively similar to that obtained with our method for $\gamma^\ast(k_1) \gamma^\ast(k_2) \to \pi^0$ \cite{Maris:2002mz, Holl:2005vu}.  However, there are currently no such discriminating data for the $\pi^0$ transition and no related computations for $M\neq \pi^0$.


\acknowledgments
%
%
We are grateful to:
V.~Braun for constructive comments made during the \emph{Workshop on Mapping Parton Distribution Amplitudes and Functions}, September 2018, European Centre for Theoretical Studies in Nuclear Physics and Related Areas (ECT*), Trento, Italy;
ECT* and its resources during that and the following \emph{Workshop on Emergent mass and its consequences in the Standard Model};
and J.~Rodr{\'{\i}}guez-Quintero for valuable interactions during the closing stages of this project.
Work supported by:
CIC (UMSNH) and CONACyT (M{\'e}xico) Grant nos.\ 4.10 and CB-2014-22117;
CONACyT (M{\'e}xico) Project ``Foins 296 - 2016'' FC (Frontiers of Science);
Center of Advanced Studies in Physics, Mathematics and Computation (CEAFMC), University of Huelva, Spain;
the Chinese Government's Thousand Talents Plan for Young Professionals;
the Chinese Ministry of Education, under the \emph{International Distinguished Professor} programme;
and U.S.\ Department of Energy, Office of Science, Office of Nuclear Physics, under contract no.~DE-AC02-06CH11357.

\appendix
\setcounter{equation}{0}
\setcounter{table}{0}
\renewcommand{\theequation}{\Alph{section}\arabic{equation}}
\renewcommand{\thetable}{\Alph{section}\arabic{table}}

\section{Topological Charge}
\label{AppendixTopology}
Beginning with the axial-vector Ward-Green-Takahashi identities, including that which is anomalous, general mass formulae for the neutral pseudoscalar mesons were derived in Ref.\,\cite{Bhagwat:2007ha}.  In the isospin-symmetry limit, those for $M=\eta, \eta^\prime$ are
\begin{align}
m_M^2 \left[
\begin{array}{c}
f^8_M \\
f^0_M
\end{array}\right]
& =
\left[
\begin{array}{c}
0 \\
n_M
\end{array}\right]
+
\left[
\begin{array}{cc}
\tfrac{1}{3} m_{12} & \tfrac{\surd 2}{3} m_{1-1} \\[1ex]
\tfrac{\surd 2}{3} m_{1-1} & \tfrac{1}{3} m_{21}
\end{array}
\right]
\left[
\begin{array}{c}
\rho^8_M \\
\rho^0_M
\end{array}\right]
\label{AnomalousMass}
\end{align}
where $m_M$ are the meson masses; $f_{M}^{8,0}$ are the octet-singlet leptonic decay constants discussed in Eqs.\,\eqref{OSmixing}, \eqref{OSfvalues}; $m_{\alpha \beta } = 2(\alpha m_l + \beta m_s)$;
\begin{equation}
n_M = \sqrt{\tfrac{3}{2}} \nu_M\,,\;
\nu_M =  \langle 0 | {\mathpzc Q} =i \frac{\alpha_S}{16\pi} \tilde F^a_{\mu\nu}F^a_{\mu\nu} | M\rangle\,,
\end{equation}
with $F^a_{\mu\nu}$ being the gluon field-strength tensor and ${\mathpzc Q}$, therefore, the topological charge density operator; and $\rho_{M}^{8,0}$ are kindred to $f_{M}^{8,0}$, \emph{viz}.\ pseudoscalar projections of the Bethe-Salpeter amplitudes onto the origin in configuration space.

Since we favoured the quark flavour basis, Sec.\,\ref{SecFlavourBasis}, then to obtain $\rho_{M}^{8,0}$ we first compute
\begin{align}
\label{eqrhoM}
i \rho_M^{l,s} & = Z_4\, {\rm tr}\int_{dk}^\Lambda \gamma_5 \chi_{M}^{l,s}(k;P)\,,
\end{align}
where $Z_4$ is the Lagrangian-mass renormalisation constant evaluated in the chiral limit, and obtain (in GeV$^2$ at $\zeta_2$)
\begin{equation}
\begin{array}{cccc}
\rho_\eta^l & \rho_\eta^s & \rho_{\eta^\prime}^l & \rho_{\eta^\prime}^s\\
0.38^2 & -0.45^2 & 0.44^2 & 0.55^2
\end{array}\,.
\end{equation}
For comparison,
$\rho_\pi^{\zeta_2} = (0.41\,{\rm GeV})^2$.
Using an obvious analogue of Eq.\,\eqref{PDAOSbasis}, these values translate to (again, GeV$^2$)
\begin{equation}
\begin{array}{cccc}
\rho_\eta^8 & \rho_\eta^0 & \rho_{\eta^\prime}^8 & \rho_{\eta^\prime}^0\\
0.50^2 & 0.033^2 & -0.37^2 & 0.57^2
\end{array}\,.
\end{equation}
Subsequently, adapting Eq.\,\eqref{OSmixing} to the present case, one finds
\begin{equation}
\label{OSrhovalues}
\begin{array}{ll}
\rho_8 =  1.7 \rho_\pi\,, & \theta_8^\rho = -28^\circ, \\
\rho_0 = 2.0 \rho_\pi\,, & \theta_0^\rho = -0.19^\circ.
\end{array}
\end{equation}
It follows from the relevant axial-vector Ward-Green-Takahashi identities that the mixing angles defined this way do not need to match those in Eq.\,\eqref{OSfvalues}; but it is supportive for the usual understanding of mixing that they are qualitatively equivalent.

One can now compute the octet-singlet in-hadron condensates \cite{Brodsky:2012ku}:
\begin{subequations}
\begin{align}
\kappa_8 & = f_8 \rho_8 =(0.33\,{\rm GeV})^3 = 2.3\,\kappa_\pi\,,\\
%
\kappa_0 & = f_0 \rho_0 =(0.34\,{\rm GeV})^3 =2.5\,\kappa_\pi\,,
\end{align}
\end{subequations}
$\kappa_\pi = (0.25\,{\rm GeV})^3$.
For any given system, the in-hadron condensate measures the coherent sum of emergent and Higgs mass generation, \emph{i.e}.\ the nonperturbatively combined influence of dynamical and explicit chiral symmetry breaking.  Notably, however, since the light-quark current-mass is very small, DCSB is overwhelmingly responsible for the size of $\kappa_\pi$, which may therefore be used to benchmark the scale of emergent mass generation.  In this connection, our computed value of
$\kappa_K = 1.5\,\kappa_\pi$
indicates that while Higgs-mass effects are noticeable, emergent mass is still dominant in the kaon, whose flavour content is $l\bar s$ or $\bar l s$.  On the other hand, if the non-Abelian anomaly is suppressed so that the Bethe-Salpeter equations produce ideally-mixed pseudoscalar bound-states, then one finds
\begin{equation}
\label{kappassbar}
\kappa_{s\bar s} = 2.2\,\kappa_\pi\,.
\end{equation}
Evidently, like the DAs in Fig.\ref{figPDAs} and the radii in Fig.\,\ref{FigRadii}, using the in-hadron condensate, one also finds that the $s$-quark defines a boundary: emergent mass generation dominates for $\hat m < \hat m_s$, but the Higgs-mass prevails on $\hat m \gtrsim \hat m_s$.

At this point, using the current-quark masses in Eq.\,\eqref{z2cqmasses}, our results for $m_{\eta,\eta^\prime}$, $f_M^{8,0}$, and Eqs.\,\eqref{AnomalousMass}, we find
\begin{equation}
\nu_\eta = (0.29\,{\rm GeV})^3\,,\;
\nu_{\eta^\prime} = (0.37\,{\rm GeV})^3;
\end{equation}
and hence the topological charge content of the $\eta^\prime$ is 2.1-times that of the $\eta$.
These results are commensurate with those obtained using a variety of other methods, \emph{e.g}., drawing from Table I in Ref.\,\cite{Singh:2013oya}:
$\nu_\eta = (0.28(2){\rm GeV})^3$,
$\nu_{\eta^\prime} = (0.36(3){\rm GeV})^3$,
$\nu_{\eta^\prime}/\nu_\eta=2.1(4)$.


\section{Interpolating Functions for Propagators and Bethe-Salpeter Amplitudes}
\label{AppendixA}
For the quark propagator, we write \cite{Bhagwat:2002tx}
\begin{subequations}
\begin{align}
S_{\mathsf f}(k) & =- i \gamma\cdot k \, \sigma_V^{\mathsf f}(k^2) + \sigma_S^{\mathsf f}(k^2)\,, \\
& = \sum_{j=1}^{j_m}\bigg[ \frac{z_j^{\mathsf f}}{i \gamma\cdot k + m_j^{\mathsf f}}+\frac{z_j^{{\mathsf f}\ast}}{i \gamma \cdot k + m_j^{{\mathsf f}\ast}}\bigg], \label{Spfit}
\end{align}
\end{subequations}
with $\Im m_j^{\mathsf f} \neq 0$ $\forall j,{\mathsf f}$.  Hence, $\sigma_{V,S}$ are meromorphic functions with no poles on the real $k^2$-axis, a feature consistent with confinement \cite{Chang:2011vu, Bashir:2012fs, Roberts:2015lja, Horn:2016rip}.  Typically, $j_m=2$ is sufficient to provide a pointwise accurate interpolation of the numerical solutions to Eq.\,\eqref{EqGap} (see, \emph{e.g}.\ Ref.\,\cite{Chang:2013pq}, Fig.\,1, and Ref.\,\cite{Shi:2015esa}, Fig.\,1).  That is also true herein and we list the interpolation parameters in Table~\ref{Table:parameters}.

\begin{table}[t]
\caption{Representation parameters. \emph{Upper panel}: Eq.\,\protect\eqref{Spfit} -- the pair $(x,y)$ represents the complex number $x+ i y$.
\emph{Lower panel}: Eqs.\,\protect\eqref{Fifit}--\protect\eqref{rhonu}.
In all cases, $a=2.75$; and $\mathpzc{l}^{\rm u}_{g_{1,2}}=1.1$, $\mathpzc{l}^{\rm u}_{g_{3}}=2.2$.
Also, $g_2$ has dimension $1/{\rm GeV}$ and $g_3$, $1/{\rm GeV}^3$.  Consequently, the listed values of $c_{g_2}$ should each be divided by the correlated value of $\Lambda_{g_{2}}^{\rm i}$ and each $c_{g_3}$ by $[\Lambda_{g_{3}}^{\rm i}]^3$.  $\Lambda^{\rm i}$ is listed in GeV.
%
%
\label{Table:parameters}
}
\begin{center}
%

\begin{tabular*}
{\hsize}
{
c@{\extracolsep{0ptplus1fil}}
c@{\extracolsep{0ptplus1fil}}
c@{\extracolsep{0ptplus1fil}}
c@{\extracolsep{0ptplus1fil}}
c@{\extracolsep{0ptplus1fil}}}\hline
${\mathsf f}$  & $z_1$ & $m_1$  & $z_2$ & $m_2$ \\\hline

$l$ & $(0.37,0.32)$ & $(0.52,0.29)$ & $(0.12,0.11)$ & $(-1.31,-0.90)$ \\
$s$ & $(0.41,0.32)$ & $(0.74,0.39)$ & $(0.12,0.10)$ & $(-1.57,-0.95)$ \\\hline
\end{tabular*}

\medskip

\begin{tabular*}
{\hsize}
{
l@{\extracolsep{0ptplus1fil}}
c@{\extracolsep{0ptplus1fil}}
c@{\extracolsep{0ptplus1fil}}
c@{\extracolsep{0ptplus1fil}}
c@{\extracolsep{0ptplus1fil}}}\hline
    & $c^{\rm i}$ & $c^{u}$ & $\phantom{-}\nu^{\rm i}$ & $\Lambda^{\rm i}$ \\\hline
$g_{1\eta}^l$
& $\phantom{-}0.94$ & $\phantom{-}0.06\phantom{6}$ & $-0.60$ & $1.35$   \\
$g_{2\eta}^l$ & $\phantom{-}0.65$ & $\phantom{-}0.006$ & $\phantom{-}3.60$ & 1.07 \\
$g_{3\eta}^l$  & $\phantom{-}0.48$ & $\phantom{-}0.04\phantom{6}$ & $\phantom{-}0.10$ & 1.10 \\\hline
$g_{1\eta}^s$ & $-2.12$ & $-0.12\phantom{6}$ & $-0.40$ & $1.35$  \\
$g_{2\eta}^s$ & $-0.94$ & $-0.01\phantom{6}$ & $\phantom{-}1.20$ & 1.18  \\
$g_{3\eta}^s$  & $-0.48$ & $-0.09\phantom{6}$ & $\phantom{-}0.10$ & 1.30\\\hline
$g_{1\eta^\prime}^l$ & $\phantom{-}0.93$ & $\phantom{-}0.07\phantom{6}$ & $-0.40$ & $1.30$  \\
$g_{2\eta^\prime}^l$ & $\phantom{-}0.72$ & $\phantom{-}0.008$ & $\phantom{-}0.40$ & 1.12  \\\hline
$g_{1\eta^\prime}^s$ & $\phantom{-}1.94$ & $\phantom{-}0.19\phantom{6}$ & $-0.22$ & $1.53$  \\
$g_{2\eta^\prime}^s$ & $\phantom{-}1.12$ & $\phantom{-}0.03\phantom{6}$ & $\phantom{-}1.60$ & 1.30  \\\hline
\end{tabular*}
\end{center}

\vspace*{-4ex}

\end{table}

Turning now to the Bethe-Salpeter amplitudes in Eq.\,\eqref{EqGenBSA}.  There are four independent scalar functions.  However, in all cases, $g_4(k;P)$ is uniformly small and is therefore neglected, as is usual \cite{Chang:2013pq, Chang:2013nia, Raya:2015gva, Raya:2016yuj}.  The same statements hold for $g_{3\eta^\prime}^{l,s}$.  We represent the remaining functions, ${\cal F}=g_{1,2,3}$, as a sum of two terms:
\begin{equation}
\label{Gpifit}
{\cal F}(k;P) = {\cal F}^{\rm i}(k;P) + {\cal F}^{\rm u}(k;P) \,,
\end{equation}
where that describing the infrared behaviour, labelled ``i'', is expressed via the following PTIR:
\begin{align}
\nonumber {\cal F}^{\rm i}(k;P) & = c_{\cal F}^{\rm i}\int_{-1}^1 \! dz \, \rho_{\nu^{\rm i}_{\cal F}}(z)\nonumber\\
 & \quad \times \bigg[ a \hat\Delta_{\Lambda^{\rm i}_{{\cal F}}}^4(k_z^2)
+ (1-a) \hat\Delta_{\Lambda^{\rm i}_{\cal F}}^5(k_z^2)
\bigg], \label{Fifit}
\end{align}
and the ultraviolet ``u'' term is expressed analogously:
{\allowdisplaybreaks
\begin{eqnarray}
\label{EFit}
{\cal F}^{\rm u}(k;P) & = & c_{\cal F}^{\rm u} \int_{-1}^1 \! dz \, \rho_{1}(z)\,
 \hat \Delta^{\mathpzc{l}^u_{\cal F}}_{1}(k_z^2)\,.
%
%
%
\end{eqnarray}
Here,
\begin{align}
\rho_\nu(z) & = \frac{\Gamma(\tfrac{3}{2}+\nu)}{\sqrt{\pi}\,\Gamma(1+\nu)}\,(1-z^2)^\nu\,, \label{rhonu}
\end{align}
$\hat \Delta_\Lambda(s) = \Lambda^2 \Delta_\Lambda(s)=\Lambda^2/[s+\Lambda^2]$,
$k_z=k-(1-z) P/2$.
%
The interpolation parameters for these scalar functions, listed in Table~\ref{Table:parameters}, were obtained by fitting their low-order Chebyshev moments:
\begin{align}
\mathcal{F}_n(k^2)=\frac{2}{\pi}\int^1_{-1}dx\sqrt{1-x^2}\,U_n(x)\,\mathcal{F}(k;P)\,,
\end{align}
$n=0,2$, $x=k\cdot P/\sqrt{k^2 P^2}$, where $U_n$ is a Chebyshev polynomial of the second kind.


\end{document}